\documentclass{aa}
\usepackage{times}
\usepackage{graphics}
\usepackage[dvips]{epsfig}

\begin{document}
\thesaurus{02(12.07.1,12.04.1,08.02.3,08.12.2,08.16.2)}
\title{Galactic microlensing with rotating binaries}
\author{M. Dominik}
\institute{Institut f\"ur Physik, Universit\"at Dortmund, D-44221 Dortmund, Germany}
\date{Received ; accepted}
\maketitle
\begin{abstract}
The influence of rotating binary systems on the light curves of galactic 
microlensing events is studied. Three different rotating binary systems
are discussed: a rotating binary lens, a rotating binary source, and the
earth's motion around the sun (parallax effect). The most dramatic effects
arise from the motion of a binary lens because of the changes of the caustic 
structure with time. 
I discuss when the treatment of a microlensing event
with a static binary model is appropriate.
It is shown that additional constraints on the unknown physical 
quantities of the lens system arise from a fit with a rotating binary lens
as well as from the earth-around-sun motion. 
For the DUO\#2 event, a fit with a rotating binary lens is presented.

\end{abstract}

\section{Introduction}
It is a fact that mass objects in a 
binary system exhibit a rotation around their center of mass. However, 
this motion has mostly been neglected in discussions of binary
sources and lenses in the context of galactic microlensing. Binary motion
can play a role through a binary lens, a binary source, and the motion
of the earth (with the observer on it) around the sun (parallax effect).
Griest \& Hu (\cite{griesthu}) have presented an examplary light curve for a rotating
binary source and have claimed that such events are rare. As we will see in
this paper, however, the influence of the rotation effects increases with
the event timescale. In addition, for the unknown halo population nothing is
known about the distribution of parameters for hypothetical binary systems.
Taking into account the large uncertainties in the position of the lens and
the relative velocity between source, lens, and observer, the rotation effects
cannot be neglected a-priori and fits with static models should be checked
for consistency. In fact, an event showing the motion of the earth around the
sun has already been detected (Alcock et. al. \cite{alcock}) and the EROS\#2 event can
be explained by microlensing of an eclipsing binary (an even more special case
of a rotating binary source) (Ansari et al. \cite{ansari}). It is however doubtful, whether
the EROS\#2 event is due to microlensing at all (Paczy{\'n}ski \cite{pac2}).
In this paper I will discuss binary motion in the source, the lens, and
in the earth-sun system (at the observer), where the most dramatic effects are caused by a rotating
binary lens through the motion of the caustics.

In Sect.~2, a description of the binary motion is given, which is needed
in the following sections. Section~3 reviews some basics of galactic microlensing.
Section~4 shows the parametrization for rotating
binary lens events and some examples for light curves. In Sect.~5,
estimates are shown which help to decide whether the treatment of a binary lens as being
static is appropriate. Section~6 discusses rotating binary sources in a similar way
as for binary lenses. In Sect.~7, I also treat the earth-around-sun motion
(parallax effect) which has been noted by Gould (\cite{gould}) and observed by
MACHO (Alcock et al. \cite{alcock}). In Sect.~8, it is shown how additional information
about physical parameters follows from the parameters for a rotating binary 
lens and from the parallax effect. For completeness, it is also noted that
an additional constraint follows from the finite source size, if the physical
size of the source is known.
Section~9 finally presents a fit for the DUO\#2 event using a rotating binary
lens, which uses different parameters than the static binary fits
already mentioned (Alard et. al \cite{alard}; Dominik \cite{dominik2}).
The appendix compares the parameters defined in Sect.~7 with the treatment
of the parallax event by MACHO (Alcock et al. \cite{alcock}).

\section{Binary motion}
In order to set the notation, 
I review some properties of the dynamics of binary systems
(see eg. Landau \& Lifshitz \cite{landau}, p.~29f.) in this section.
Let us consider an object of mass $\mu_1$ at $\vec r_1$ and
an object of mass $\mu_2$ at $\vec r_2$. The Lagrangian of this
system is given by 
\begin{equation}
L = \frac{1}{2}\mu_1 {\dot {\vec r_1}}^2
+ \frac{1}{2}\mu_2 {\dot {\vec r_2}}^2 
- V\left(\left|{\vec r_1}-{\vec r_2}\right|\right)\,,
\end{equation}
where $V$ is the gravitational potential
\begin{equation}
V\left(\left|{\vec r_1}-{\vec r_2}\right|\right)
= -G \frac{\mu_1 \mu_2}{\left|{\vec r_1}-{\vec r_2}\right|}\,.
\end{equation}
This is the Kepler problem.

Let $\vec r$ be the difference vector,
$\vec R$
the coordinate of the center of mass,
$M$ the total mass and $\mu$ the reduced mass
given by
\begin{eqnarray}
\vec r & = & {\vec r_1}- {\vec r_2}\,, \\
\vec R & = & \frac{{\mu_1 {\vec r_1} + \mu_2 {\vec r_2}}}{\mu_1 + \mu_2}\,, \\
M & = & \mu_1+\mu_2\,, \\
\mu & = & \frac{\mu_1 \mu_2}{\mu_1 + \mu_2}\,. 
\end{eqnarray}
With these definitions, the Lagrangian can be written as
\begin{equation}
L = \frac{1}{2} M {\dot {\vec R}\,}^2 + \frac{1}{2} \mu {\dot {\vec r}\,}^2
- V(|{\vec r}\,|)\,.
\end{equation}
The Euler-Lagrange equation for $\vec R$ is
\begin{equation}
0 = \frac{d}{dt}\,\frac{\partial L}{\partial {\dot R_j}} =
\frac{d}{dt}\,\left(M {\dot R_j}\right)\,,
\end{equation}
so that $\dot{\vec R} = \mbox{const.}$, i.e. the center of mass moves
uniformly. If one chooses a coordinate system with the origin at the
center of mass, one has $\vec R = 0$ and therefore
\begin{equation}
{\vec r_1} = \frac{\mu_2}{\mu_1+\mu_2}\,{\vec r}\quad \mbox{and}\quad
{\vec r_2} = -\frac{\mu_1}{\mu_1+\mu_2}\,{\vec r}\,. \label{commot}
\end{equation}
As can be seen from the Lagrangian, $\vec r(t)$ gives the motion of
a particle of mass $\mu$ in the gravitational potential
\begin{equation}
V(|\vec r\,|) = -G\, \frac{\mu M}{|\vec r\,|\,}\,.
\end{equation}

For a gravitationally bound system, the trajectory
is an ellipse in a plane perpendicular to the angular momentum $\vec L$, 
where the origin (center of mass) is in a focus of the 
ellipse. Let $\varepsilon$ be the eccentricity and $a$ the semimajor axis.
With polar coordinates $(r,\varphi)$, the trajectory is given by
\begin{equation}
r(\varphi) = \frac{q}{1+\varepsilon \cos \varphi}\,,
\end{equation}
where $q = a(1-\varepsilon^2)$. 
The minimal value is $r_\mathrm{min} = a(1-\varepsilon)$ obtained for $\varphi = 0$,
and the maximal value is $r_\mathrm{max} = a(1+\varepsilon)$ obtained for
$\varphi = \pi$.

Therefore, one can parametrize the curve with a parameter $\xi$ as
\begin{equation}
r(\xi) = a(1-\varepsilon \cos \xi)\,.
\end{equation}
The components along the semimajor axis
($x$-direction) and the semiminor axis ($y$-direction) follow as
\begin{eqnarray}
x(\xi) &=& a(\cos \xi - \varepsilon)\,, \label{solkep1} \\
y(\xi) &=& a\sqrt{1-\varepsilon^2} \sin \xi\,. \label{solkep2}  
\end{eqnarray}

The time dependence is given by 
\begin{equation}
t = \sqrt{\frac{a^3}{GM}}\,(\xi - \varepsilon \sin \xi)\,,
\end{equation}
so that $t=0$ corresponds to the point $(r_\mathrm{min},0)$.
One sees that in general this equation cannot be solved analytically 
for $\xi$ to yield $x(t)$ and $y(t)$ but instead has to be solved numerically.
Changing $\xi$ to $\xi+2\pi$ corresponds to one revolution, so that the
period is given by
\begin{equation}
T = 2\pi \,\sqrt{\frac{a^3}{GM}}\,.
\end{equation}

Since, for $n \in \bbbz$, $x(\xi+n 2\pi) = x(\xi)$ and $y(\xi+n 2\pi) = y(\xi)$ one can
subtract full periods $n$ from the given time $t$ and solve
\begin{equation}
t' = t - nT = \sqrt{\frac{a^3}{GM}}\,(\xi - \varepsilon \sin \xi)
\label{xieq}
\end{equation}
for a $\xi \in [0,2\pi)$. 
With  
\begin{equation}
\lfloor x \rfloor = k\,\quad \mbox{with}\quad k \in \bbbz\,, k \leq x < k+1
\end{equation}
and the period $T$, Eq.~(\ref{xieq}) can be written as
\begin{equation}
2\pi\left(\frac{t}{T}-\left\lfloor \frac{t}{T} \right\rfloor \right) = 
\xi-\varepsilon \sin \xi\,.
\end{equation}
For $\xi = 0$ at $t=t_0$ one has
\begin{equation}
2\pi\left(\frac{t-t_0}{T}-\left\lfloor \frac{t-t_0}{T}\right\rfloor\right) = 
\xi-\varepsilon \sin \xi\,.
\end{equation}
For $\xi = \xi_0$ at $t = t_1$, 
$\xi = 0$ is obtained for 
\begin{equation}
t_0 = t_1 - \frac{T}{2\pi}(\xi_0 - \varepsilon \sin \xi_0)\,,
\end{equation}
so that
\begin{eqnarray}
2\pi\left(\frac{t-t_1}{T}-\left\lfloor\frac{t-t_1}{T}
+\frac{1}{2\pi}(\xi_0-\varepsilon\sin\xi_0)\right\rfloor\right)\,+   \nonumber \\ 
+\,\xi_0-\varepsilon \sin
\xi_0 = 
\xi-\varepsilon \sin \xi   
\label{getxi}
\end{eqnarray}
will yield a $\xi \in [0,2\pi)$.

Let $r = |\vec r\,|$ and $v = |\dot{\vec r}\,|$.
The total energy $E$, which is the sum of the kinetic energy $T$ and
the potential energy $V$ 
\begin{equation}
E = T + V = \frac{1}{2}\mu v^2 - \frac{G\mu M}{r}\,, \label{en1}
\end{equation}
is related to the semimajor axis $a$ by
\begin{equation}
E = - \frac{G\mu M}{2a}\,. \label{en2}
\end{equation}
From Eq.~(\ref{en1}) and Eq.~(\ref{en2}), and with
\begin{equation}
v_\mathrm{circ} = \frac{2\pi}{T}\,a\,,
\end{equation}
the velocity as a function of $r$ reads
\begin{equation}
v(r) = v_\mathrm{circ}\,\sqrt{\frac{2a}{r}-1}\,.
\end{equation}
The maximal velocity is obtained for $r_\mathrm{min}$ as
\begin{equation}
v_\mathrm{max} = v(r_\mathrm{min}) = v_\mathrm{circ}\,\sqrt{\frac{1+\varepsilon}{1-\varepsilon}}\,,
\end{equation}
and the minimal velocity is obtained for $r_\mathrm{max}$ as
\begin{equation}
v_\mathrm{min} = v(r_\mathrm{max}) = v_\mathrm{circ}\,\sqrt{\frac{1-\varepsilon}{1+\varepsilon}}\,,
\end{equation}
The ratio between $v_\mathrm{max}$ and $v_\mathrm{min}$ is
\begin{equation}
\rho_v = \frac{v_\mathrm{max}}{v_\mathrm{min}} = \frac{1+\varepsilon}{1-\varepsilon}\,.
\end{equation}
Values for $\rho_v$ for different eccentricities $\varepsilon$ are shown
in Table~\ref{rhovtab}.
From the virial theorem, one obtains for the expectation values of the 
kinetic and the potential energy the relation
\begin{equation}
2\,<\!\!T\!\!> = - <\!\!V\!\!>\,,
\end{equation}
so that one obtains for the radius $r$ and the velocity $v$ the relations
\begin{equation}
<\!\!\frac{1}{r}\!\!> = \frac{1}{a}\,,
\end{equation}
and
\begin{equation}
<\!\!v^2\!\!> = \frac{GM}{a} = \frac{4\pi^2}{T^2}\,a^2 = v_\mathrm{circ}^2\,.
\end{equation}

\begin{table}[htbp]
\caption{The ratio $\rho_v$ between the maximal and the minimal velocity for
different eccentricities $\varepsilon$}
\begin{flushleft}
\begin{tabular}{cc}
\hline\noalign{\smallskip}
$\varepsilon$ & $\rho_v$ \\
\noalign{\smallskip}\hline\noalign{\smallskip} 
0 & 1.000 \\ 
0.01 & 1.020 \\ 
0.05 & 1.105 \\ 
0.1 & 1.111 \\ 
0.2 & 1.500 \\ 
0.3 & 1.857 \\ 
0.4 & 2.333 \\ 
0.5 & 3.000 \\ 
0.6 & 4.000 \\ 
0.7 & 5.667 \\ 
0.8 & 9.000 \\ 
0.9 & 19.000 \\
\noalign{\smallskip}\hline
\end{tabular}
\end{flushleft}
\label{rhovtab}
\end{table}

\section{Some basics of gravitational lensing}
The effect of light bending by a point mass $M$ at the distance $D_\mathrm{d}$ 
from the observer and at distance $D_\mathrm{ds}$ from the source object
which is located at a distance $D_\mathrm{s}$ from the observer 
can be described by the
gravitational lens equation
\begin{equation}
\vec y = \vec x - \frac{\vec x}{|\vec x|^2}\,,
\end{equation}
(e.g. Schneider et al. \cite{sef}), 
where $\vec y$ is a dimensionless coordinate in the plane perpendicular to
the line-of-sight observer-lens at the position of the source (source plane)
and $\vec x$ is a dimensionless coordinate in a corresponding plane 
at the position of the lens (lens plane).
The physical position of the light ray connecting the source and the observer
is given by 
\begin{equation}
\vec \xi = r_\mathrm{E}\,\vec x
\end{equation}
in the lens plane 
and
\begin{equation}
\vec \eta = \frac{D_\mathrm{s}}{D_\mathrm{d}}\,r_\mathrm{E}\,\vec y
= r_\mathrm{E}'\,\vec y\,.
\end{equation}
In these equations, $r_\mathrm{E}$ denotes the Einstein radius, given by
\begin{equation}
r_\mathrm{E} = \sqrt{\frac{4GM}{c^2}\,\frac{D_\mathrm{d}\,D_\mathrm{ds}}{D_\mathrm{s}}}\,,
\label{Einsteinradius}
\end{equation}
and $r_\mathrm{E}' = \frac{D_\mathrm{s}}{D_\mathrm{d}}\,r_\mathrm{E}$ 
denotes the projected Einstein radius.
For a system of $N$ lenses at positions $\vec x^{\,(i)}$ with mass fractions
$m_i$, the lens equation reads
\begin{equation}
\vec y = \vec x - \sum\limits_{i=1}^{N}\,\frac{\vec x - \vec x^{\,(i)}}
{\left|\vec x - \vec x^{\,(i)}\right|^2}\,.
\end{equation}

Let $v_{\perp}$ be the velocity of the relative motion between lens, source,
and observer as measured in the lens plane. If one considers a coordinate
system in which the observer and the source are at rest, $v_{\perp}$ gives
the relative motion of the lens. Alternatively, one can consider a coordinate
system in which the observer and the lens are at rest, so that $v_{\perp}$
gives the motion of the source position as projected onto the lens plane.
In either case, a characteristic timescale of the motion (and therefore
of the event) is given by
\begin{equation}
t_\mathrm{E} = \frac{r_\mathrm{E}}{v_{\perp}}\,.
\end{equation}
This definition means that the moving object transverses one Einstein radius in
the lens plane in the time $t_\mathrm{E}$. 

Let $t_\mathrm{max}$ denote the time at the closest approach to the line-of-sight 
and $u_\mathrm{min}$ the impact parameter at $t_\mathrm{max}$ in units of the 
Einstein radius $r_\mathrm{E}$. For a point source and a point-mass lens, one
obtains for the impact parameter at time $t$
\begin{equation}
u(t) = \sqrt{u_\mathrm{min}^2 + [p(t)]^2}
\end{equation}
with
\begin{equation}
p(t) = \frac{t-t_\mathrm{max}}{t_\mathrm{E}}\,,
\end{equation}
and the amplification is given by (e.g. Paczy{\'n}ski \cite{pac1})
\begin{equation}
A(u) = \frac{u^2 +2 }{u\sqrt{u^2+4}}\,.
\end{equation}
The light curve for an event involving a point source and a point-mass lens
is therefore described by the 3 parameters $t_\mathrm{E}$, $t_\mathrm{max}$,
and $u_\mathrm{min}$.

\section{Rotating binary lenses}
\label{sec:robin}
For a rotating binary lens,
one needs the projection of the trajectory onto
the lens plane. The orientation of the rotating system relative to 
the lens plane is given by two angles
$\beta$ and $\gamma$. For $\beta = 0$ and $\gamma = 0$, $x$ is chosen along
$x_1$, $y$ along $x_2$ and the angular momentum $\vec L$ is towards the
observer ($x_3$-direction). The angle $\beta$ describes a rotation of the
lens system around $x_1$ and the angle 
$\gamma$ a following rotation of the lens system
around $x_2$. This means that one has the transformation
\begin{equation}
\left(\begin{array}{c} x_1 \\ x_2 \\ x_3 \end{array}\right) 
=  \frac{1}{r_\mathrm{E}}\,\mathcal{R}_1\,
\left( \begin{array}{c} x \\ y \\ z\end{array}\right)\,
\end{equation}
with
\begin{equation}
\mathcal{R}_1 = \left(\begin{array}{ccc} \cos \gamma & \sin \beta \sin \gamma &
\cos \beta \sin \gamma \\
0 & \cos \beta & - \sin \beta \\
-\sin \gamma & \sin \beta \cos \gamma & \cos \beta \cos \gamma 
\end{array}\right)\,.
\end{equation}
Since $z = 0$ and the $x_3$-value is redundant, the transformation reduces
to
\begin{equation}
\left(\begin{array}{c} x_1 \\ x_2 \end{array}\right)
= \frac{1}{r_\mathrm{E}}\,\left(\begin{array}{cc} \cos \gamma & \sin \beta \sin \gamma \\
0 & \cos \beta 
\end{array}\right) \left( \begin{array}{c} x \\ y \end{array}\right)\,.
\label{blang}
\end{equation}
A rotation around $x_3$ is not considered here, since it can be put into 
the orientation $\alpha$ of the source trajectory.

Altogether, one needs the following parameters for lensing by
a rotating binary lens:
\begin{itemize}
\item The point of time $t_b$ of the closest approach of the 
source to the center of mass of the lens system,
\item the characteristic time $t_\mathrm{E} = r_\mathrm{E}/v_{\perp}$,
\item the mimimal projected distance $b$ in the lens plane between source 
and center of mass of the lens system in units of the Einstein radius,
\item the angle $\alpha$ between the $x_1$-direction and the direction of the 
	projected source trajectory,
\item the mass fraction $m_1 = \mu_1/M$,
\item the semimajor axis in units of the Einstein radius $\rho = a/r_\mathrm{E}$,
\item the rotation angle $\beta$,
\item the rotation angle $\gamma$,
\item the period $T$,
\item the eccentricity $\varepsilon$,
\item the phase $\xi_0$ at $t = t_b$.
\end{itemize}
Compared with the static binary lens, one needs 5 additional parameters.

The position of a hypothetical object of mass $\mu$ is therefore,
using Eqs.~(\ref{solkep1}),~(\ref{solkep2}), and~(\ref{blang}),
\begin{eqnarray}
x_1(t) & = & \rho \Big[\cos \gamma(\cos \xi(t)-
\varepsilon)
\,+ \nonumber \\
 & & + \, \left.\sin \beta \sin \gamma \sqrt{1-\varepsilon^2}\sin \xi(t)\right]\,, \\
x_2(t) & = & \rho \cos \beta \sqrt{1-\varepsilon^2}\sin \xi(t)\,,
\end{eqnarray}
and, with Eq.~(\ref{commot}), 
the positions of the masses $\mu_1$ and $\mu_2$ are
\begin{equation}
{\vec x}^{\,(1)}(t) = (1-m_1) {\vec x}(t)\,, \quad
{\vec x}^{\,(2)}(t) = -m_1 {\vec x}(t)\,.
\end{equation}
From Eq.~(\ref{getxi}), the 
value of $\xi \in [0,2\pi)$ for a given $t$ is given by
\begin{eqnarray}
2\pi\left(\frac{t - t_b}{T} - 
\left\lfloor\frac{t - t_b}{T} + \frac{1}{2\pi} 
\left(\xi_0 - \varepsilon \sin \xi_0\right)\right\rfloor\right)\,+  \nonumber
   \\ +\,\xi_0 - 
\varepsilon\sin \xi_0
= \xi - \varepsilon \sin \xi\,. 
\end{eqnarray}

Examples for rotating binary lenses are shown in Figs.~\ref{RBXT}
and~\ref{RBT}. For both figures, $\beta = \gamma = 
\varepsilon = \xi_0 = 0$, and $\rho$ has been chosen as $2\chi$.
For Fig.~\ref{RBXT}, the lens model
BL for the MACHO LMC\#1 event has been used (Dominik \& Hirshfeld \cite{domhirsh}),
so that $t_\mathrm{E} = 16.27~d$, $t_b = 433.18$,
$\rho = 0.4077$, $m_1 = 0.463$, $\alpha = 1.151$, $b = 0.146$,
and $m_\mathrm{base} = 4.5170$. 
$m_\mathrm{base}$ denotes the negative observed magnitude at the unlensed state and
$f$ the contribution of the source to the total light at unlensed state.
The rotation period has been chosen as $T$ = 365~d, 
100~d, 50~d, and 25~d.
For Fig.~\ref{RBT}, the lens model
BL0 for the OGLE\#7 event has been used (Dominik \cite{dominik2})\footnote{This model
coincides with that of Alard et. al (\cite{alard}).},
so that $t_\mathrm{E} = 80.88~d$, $t_b = 1173.25$,
$\rho = 1.131$, $m_1 = 0.506$, $\alpha = 2.297$, $b = 0.048$,
$f = 0.557$,
and $m_\mathrm{base} = -17.5171$. 
The rotation period has been chosen as $T$ = 3000~d, 
1000~d, 365~d, and 100~d.  

One sees that dramatic effects occur if the period is small, especially
if additional caustic crossings occur. But even for a period of 365~days,
a deviation from the MACHO LMC\#1-fit is visible, and for a period of 1000~days a
second peak for parameters near the OGLE\#7-fit occurs. This constellation
looks a little like the DUO\#2 event. A corresponding model is discussed
in Sect.~\ref{DUO2robinfit}. Note that the rotation period is about 12 times larger
than the timescale $t_\mathrm{E}$ for this constellation, nevertheless a dramatic 
effect occurs in the light curve. Note also that the example used by 
Griest \& Hu (\cite{griesthu}) for 
rotating binary
sources used a period $T$ which is 4 times smaller than $t_\mathrm{E}$.

\begin{figure*}
\vspace{10.5cm}
\caption{Rotating binary lenses with parameters as for the BL-fit of MACHO LMC\#1.
$\beta = \gamma = 
\varepsilon = \xi_0 = 0$. {\bf a} $T=365$~d, {\bf b} $T=100$~d, {\bf c} $T=50$~d,
{\bf d} $T=25$~d.}
\label{RBXT}
\end{figure*}

\begin{figure*}
\vspace{10.5cm}
\caption{Rotating binary lenses with parameters as for the BL0-fit of OGLE\#7.
$\beta = \gamma = 
\varepsilon = \xi_0 = 0$. {\bf a} $T=3000$~d, {\bf b} $T=1000$~d,
{\bf c} $T=365$~d, {\bf d} $T=100$~d.}
\label{RBT}
\end{figure*}

\section{When is the rotation effect negligible?}
Let us consider a fit for a static binary lens. As discussed by
Dominik (\cite{dominik1}), cited as D97a in the following,
the rotation period $T$ can be estimated using the
timescale $t_\mathrm{E}$ and distributions of the lens position and the
velocity $v_{\perp}$. In a similar way, one can also obtain probability
distributions for the
ratio of the timescales
\begin{equation}
R_T = \frac{t_\mathrm{E}}{T}\,,
\end{equation}
and the ratio of the velocities
of the binary motion of the lens and the perpendicular motion with respect
to the line-of-sight
\begin{equation}
R_v = \frac{v_\mathrm{circ}}{v_{\perp}}\,.
\end{equation}
Note that irrespective of the eccentricity
$<\!\!v^2\!\!> = v_\mathrm{circ}^2$, and the maximal and minimal velocities
are of the same order for moderate eccentricities as shown in
Sect.~2. The projection of the velocity of the lens system
to a plane perpendicular to the line-of-sight 
may be lower than this, but the velocity is perpendicular to the line-of-sight
at least twice a period.
Expressing $R_T$ and $R_v$ in terms of the fit parameters, the lens distance
$x = D_\mathrm{d}/D_\mathrm{s}$, and the dimensionless velocity parameter
$\zeta = v_{\perp}/v_\mathrm{c}$ (where $v_\mathrm{c}$ is a characteristic velocity),
yields
\begin{equation}
R_T = \frac{c}{4\pi}\,\sqrt{\frac{t_\mathrm{E}}{\rho^3\,D_\mathrm{s}\,v_\mathrm{c}\,x(1-x)\,\zeta}}\,,
\end{equation}
and
\begin{equation}
R_v = \frac{c}{2}\,\sqrt{\frac{t_\mathrm{E}}{\rho\,D_\mathrm{s}\,v_\mathrm{c}\,x(1-x)\,\zeta}}\,.
\end{equation}
One sees that
\begin{equation}
R_v = 2\pi\rho\, R_T\,.
\end{equation}
Note that there are two problems with these quantities. First, as discussed
above, the projected trajectory of the binary components are not circles.
Second, from fits with static binary models, one only gets the projected
distance $2\chi$ between the objects, where $\rho \geq \chi$ in general and
$\rho = 2\chi$ for circular projected trajectories.

Since $R_T,R_v \propto \sqrt{t_\mathrm{E}}$ and $R_T \propto \sqrt{1/\rho^3}$,
$R_v \propto \sqrt{1/\rho}$, the rotation of binary lenses is most important
for events with long $t_\mathrm{E}$ and small $\rho$.\footnote{It is implied that
one has an effect from binarity. For $\rho \to 0$, this effect vanishes.
However, if one sees the binarity, the rotation is especially important 
for small $\rho$.}

Since $R_T$ and $R_v$ are of the form
\begin{equation}
G(t_\mathrm{E},x,\zeta) = G_0(t_\mathrm{E})\,[x(1-x)]^k\,\zeta^l
\end{equation}
with $k = l = -\frac{1}{2}$, estimates and probability distributions
can be derived using the approach presented in D97a.
Let $H(x)\,dx$ be the probability for finding $x$ in $[x,x+dx]$ (where
$H(x)$ is proportional to the mass density $\rho(x)$)\footnote{$H(x) = \rho(x)/\rho_0$,
where $\rho_0$ is an arbitrary characteristic mass density, so that at the distance
$x_0$, where $\rho(x_0) = \rho_0$, one has $H(x_0) = 1$.},
$\widetilde{K}(\zeta)\,d\zeta$ be the probability for finding $\zeta$ 
in $[\zeta,\zeta+d\zeta]$, and let
$T(r,s)$ be defined as
\begin{equation}
T(r,s)  =  \int \left[x(1-x)\right]^r\,H(x)\, 
\zeta^s\,\widetilde{K}(\zeta)\,d\zeta\,dx\,, \label{eqt}
\end{equation}
which separates as
\begin{equation}
T(r,s) = \Xi(r)\,W(s)\,,
\end{equation}
where
\begin{eqnarray}
\Xi(r) & =  & \int \left[x(1-x)\right]^r\,H(x)\,dx\,, \label{eqxi} \\ 
W(s) & = &  \int \zeta^s\,\widetilde{K}(\zeta)\,d\zeta\,, \label{eqw} 
\end{eqnarray}
if the velocity distribution does not
depend on $x$.
Following D97a, the expectation values for a quantity $G$
assuming an unknown mass
distribution are given by
\begin{equation}
<\!\!G\!\!> = G_0\,\frac{T(k+1,l)}{\Xi(1)} = G_0\,F(-1,k,l)\,.
\end{equation}

Let us adopt the simple galactic halo model of D97a with 
a velocity distribution of
\begin{equation}
\widetilde{K}(\zeta) = 2 \zeta\,\exp\left\{-\zeta^2\right\}
\end{equation}
where $v_\mathrm{c}$ has been chosen so that
$<\!\!v_{\perp}^2\!\!> = v_\mathrm{c}^2$,
and the mass density of halo objects being
\begin{equation}
\rho(r) = \rho_0\,\frac{R_\mathrm{GC}^2}{r^2}\,,
\end{equation}
where $r$ measures the distance from the Galactic center, $R_\mathrm{GC}$ is
the distance from the sun to the Galactic center, 
and $\rho_0$ is the local density at the position of the sun.
With the values 
$D_\mathrm{s} = 50~\mbox{kpc}$ and 
$R_\mathrm{GC} = 10~\mbox{kpc}$, and the halo being extended up to the LMC, located
at $82^{\circ}$ from the Galactic center as seen from the observer,
one obtains $\Xi(\frac{1}{2}) = 0.105$, $\Xi(1) = 0.0407$,
$W(-\frac{1}{2}) = 1.225$ and therefore $F(-1,-\frac{1}{2},-\frac{1}{2})
= 3.16$.
Using slightly
different values for $D_\mathrm{s}$ ($55~\mbox{kpc}$) and 
$R_\mathrm{GC}$ ($8.5~\mbox{kpc}$), and varying the 
core radius $a$ between 0 and 8~kpc yields estimates which differ
by about~5~\%.

For the expectation values, one obtains 
\begin{eqnarray}
<\!\!R_T\!\!> & = & 1.23\cdot 10^{-3}\,\rho^{-3/2}\,\left(\frac{v_\mathrm{c}}{210~\mbox{km/s}}\right)^{-1/2}\,
\,\left(\frac{t_\mathrm{E}}{1~\mbox{d}}\right)^{1/2} \\
<\!\!R_v\!\!> & = & 7.74\cdot 10^{-3}\,\rho^{-1/2}\,\left(\frac{v_\mathrm{c}}{210~\mbox{km/s}}\right)^{-1/2}\,
\,\left(\frac{t_\mathrm{E}}{1~\mbox{d}}\right)^{1/2} \label{haloestlast} 
\end{eqnarray}
The probability density for 
\begin{equation}
\kappa_R = R_T/\!\!<\!\!R_T\!\!> 
= R_v/\!\!<\!\!R_v\!\!>
\end{equation}
is given in D97a as 
\begin{eqnarray}
p(\kappa_R) = \frac{4\left[\Xi(1)\right]^3}{\left[\Xi(\frac{1}{2})\,
W(-\frac{1}{2})\right]^4}\,\kappa_R^{-5}\,
\int \frac{H(x)}{x(1-x)}\,\cdot  \nonumber \\
\cdot\,\exp\left\{-\left(\frac{1}{\kappa_R\,\sqrt{
x(1-x)}}\,\frac{\Xi(1)}{\Xi(\frac{1}{2})\,W(-\frac{1}{2})}\right)^4\right\}
\,dx\,, 
\end{eqnarray}
and the probability density for $\lambda_R = \mbox{lg}\,\kappa_R$ is given as
\begin{eqnarray}
\psi(\lambda_R) = \frac{4\,\ln 10}{10^{4 \lambda_R}}\,
\frac{\left[\Xi(1)\right]^3}{\left[\Xi(\frac{1}{2})\,
W(-\frac{1}{2})\right]^4}\,
\int \frac{H(x)}{x(1-x)}\,\cdot \nonumber \\ \cdot\,\exp\left\{-\left(\frac{1}{10^{\lambda_R}\,\sqrt{
x(1-x)}}\,\frac{\Xi(1)}{\Xi(\frac{1}{2})\,W(-\frac{1}{2})}\right)^4\right\}
\,dx\,.
\end{eqnarray}
These probability densities are shown in Fig.~\ref{halo:pkr}, where symmetric
intervals around $<\!\!R_T\!\!>$ or $<\!\!R_v\!\!>$ containing probabilities
of 68.3~\% and 95.4~\% respectively are shown for $\lambda_R$.
The bounds of these intervals are
also shown in Table~\ref{halo:lambounds}.
The smallest and the largest value in the 95.4~\%-interval differ
by a factor of about 5.

\begin{table*}[htb]
\caption[ ]{The bounds of symmetric intervals 
around 
$<\!\!R_T\!\!>$ or $<\!\!R_v\!\!>$ on a logarithmic scale 
which correspond to probabilities of 68.3~\% and
95.4~\%}
\begin{flushleft}
\begin{tabular}{cccccc}
\hline
\noalign{\smallskip}
$\Delta\lambda_{68.3}$ & $\Delta\lambda_{95.4}$ & 
$10^{-\Delta\lambda_{68.3}}$ & $10^{\Delta\lambda_{68.3}}$ & 
$10^{-\Delta\lambda_{95.4}}$ & $10^{\Delta\lambda_{95.4}}$ \\
\noalign{\smallskip}\hline\noalign{\smallskip}
0.1844 & 0.3317 & 0.654 & 1.53 & 0.466 & 2.15 \\ 
\noalign{\smallskip}\hline
\end{tabular}
\end{flushleft}
\label{halo:lambounds}
\end{table*}

\begin{figure*}
\vspace{7cm}
\caption{The probability densities {\bf a} $p(\kappa_R)$ (left)
and {\bf b} $\psi(\lambda_R)$ (right)
with symmetric 68.3~\% and 95.4~\% intervals around $<\!\!R_T\!\!>$ 
or $<\!\!R_v\!\!>$}
\label{halo:pkr}
\end{figure*} 

For the binary lens fits to MACHO LMC\#1 (Dominik \& Hirshfeld \cite{domhirsh}), 
one obtains the values shown in Table~\ref{MACHOest}.
Since the true semimajor axis $a = \rho\,r_\mathrm{E}$ 
is not yielded by the fit, the estimates refer to
$\rho = \chi$, which corresponds to a
minimal value of the period $T_\mathrm{min}$, because
$\rho \geq \chi$ for any gravitationally bound system and
$T \propto \rho^{3/2}$. The timescale $t_\mathrm{E}^{\,(2)}$ corresponds to the
Einstein radius of the smaller mass with mass fraction $1-m_1$,
\begin{equation}
t_\mathrm{E}^{\,(2)} = t_\mathrm{E}\,\sqrt{1-m_1}\,.
\end{equation}

\begin{table}[htbp]
\caption[ ]{MACHO LMC\#1: Estimates of $R_T$ and $R_v$ 
for different binary lens models.}
\begin{flushleft}
\begin{tabular}{lcccccc}
\hline\noalign{\smallskip}
 & BL & BL1 & BA & BA1 & BA2 \\ \noalign{\smallskip}\hline\noalign{\smallskip} 
\rule[-1ex]{0ex}{3.5ex}$t_\mathrm{E}$ [d] & 16.27 & 17.53 & 685 & 155 & 35.7 \\
\rule[-1ex]{0ex}{3.5ex}$t_\mathrm{E}^{\,(2)}$ [d] & --- & --- & 17.57 & 15.15 & 17.72 \\
\rule[-1ex]{0ex}{3.5ex}$<\!\!R_T\!\!>$&
0.054 & 0.050 & 0.0086 & 0.0047 & 0.0030 \\ 
\rule[-1ex]{0ex}{3.5ex}$<\!\!R_v\!\!>$&
0.069 & 0.069 & 0.13 & 0.065 & 0.034 \\ 
\noalign{\smallskip}\hline
\end{tabular}
\end{flushleft}
\label{MACHOest}
\end{table} 

One sees that the rotation is not likely to play a dominant effect, however
a marginal effect may show up. For the wide binary models (BA, BA1, BA2),
the peak
arises from the passage near the smaller mass on a timescale
$t_\mathrm{E}^{(2)}$, so that the influence from the binary rotation on the
peak will be smaller than estimated using $t_\mathrm{E}$.

\section{Rotating binary sources}
\label{rosrc}
In their discussion of binary sources, Griest \& Hu (\cite{griesthu}) have also mentioned 
their rotation. Here I show that
the parameters for a rotating binary source can be chosen in analogy to the
rotating binary lens. 
Let the relative motion of the lens perpendicular to the source-observer-line
projected to the source plane be
\begin{equation}
{\vec y}^{\,\mathrm{(lens)}}(t) = \left(\begin{array}{c} \cos \widetilde{\alpha}
\\ \sin \widetilde{\alpha} \end{array}\right)\,\frac{t-\widetilde{t_b}}{t_\mathrm{E}}\,
+ \, \left(\begin{array}{c} -\sin \widetilde{\alpha}
\\ \cos \widetilde{\alpha} \end{array}\right)\,\widetilde{b}\,,
\end{equation}
where $\widetilde{t_b}$ is the point of time of closest approach to
the center of mass of the binary source system.
The orientation of the rotating system relative to 
the source plane is given by two angles
$\widetilde{\beta}$ and $\widetilde{\gamma}$. 
For $\widetilde{\beta} = 0$ and $\widetilde{\gamma} = 0$, $x$ is chosen along
$y_1$, $y$ along $y_2$ and the angular momentum $\vec L$ is towards the
observer ($y_3$-direction). The angle $\widetilde{\beta}$ describes a rotation of the
source system around $y_1$ and the angle 
$\widetilde{\gamma}$ a following rotation of the source system
around $y_2$. 
As for the rotating binary lens, one has the transformation 
\begin{equation}
\left(\begin{array}{c} y_1 \\ y_2 \end{array}\right)
= \frac{1}{r_\mathrm{E}'}\,\left(\begin{array}{cc} \cos \widetilde{\gamma} 
& \sin \widetilde{\beta} \sin \widetilde{\gamma} \\
0 & \cos \widetilde{\beta} 
\end{array}\right) \left( \begin{array}{c} x \\ y \end{array}\right)
\,. \label{srcproj}
\end{equation}
A rotation around $y_3$ need not to be considered here, since it can be put into 
the orientation $\widetilde{\alpha}$ of the lens trajectory.

For lensing of a rotating binary source one needs the following parameters:
\begin{itemize}
\item The point of time $\widetilde{t_b}$ of the closest approach of the 
lens to the center of mass of the source system,
\item the characteristic time $t_\mathrm{E} = r_\mathrm{E}/v_{\perp}$,
\item the mimimal projected distance $\widetilde{b}$ in the source plane between lens 
and center of mass of the source system in units of the projected 
Einstein radius,
\item the angle $\widetilde{\alpha}$ between the $y_1$-direction and the direction of the 
	projected lens trajectory,
\item the luminosity offset ratio $\omega$,
\item the mass fraction $\widetilde{m}_1$ of source object 1,
\item the semimajor axis in units of the projected Einstein 
radius $\widetilde{\rho} = a/r_\mathrm{E}'$,
\item the rotation angle $\widetilde{\beta}$,
\item the rotation angle $\widetilde{\gamma}$,
\item the period $\widetilde{T}$,
\item the eccentricity $\widetilde{\varepsilon}$,
\item the phase $\widetilde{\xi_0}$ at $t = \widetilde{t_b}$.
\end{itemize}

Compared with the static binary source, one needs 6 additional parameters.

From Eqs.~(\ref{solkep1}), (\ref{solkep2}), and~(\ref{srcproj}), 
the position of a hypothetical object of the reduced mass is given by
\begin{eqnarray}
y_1(t) & = & \widetilde{\rho} \Big[\cos \widetilde{\gamma}(\cos \xi(t)-
\widetilde{\varepsilon})\,+ \nonumber \\
 &  & +\,\left.\sin \widetilde{\beta} \sin \widetilde{\gamma} \sqrt{1-\widetilde{\varepsilon}^2}\sin \xi(t)\right]\,, \\
y_2(t) & = &\widetilde{\rho} \cos \widetilde{\beta}
 \sqrt{1-\widetilde{\varepsilon}^2}\sin \xi(t)\,,
\end{eqnarray}
and the positions of the source objects 1 and 2 are, using Eq.~(\ref{commot}),
\begin{equation}
{\vec y}^{\,(1)}(t) = (1-\widetilde{m}_1)\, {\vec y}(t)\,, \quad
{\vec y}^{\,(2)}(t) = -\widetilde{m}_1\, {\vec y}(t)\,.
\end{equation}
Note that for $\widetilde{\xi_0} = 0$ and $\widetilde{\beta} =
\widetilde{\gamma} = 0$, one obtains $y_1(\widetilde{t_b}) = 
\widetilde{\rho}(1-\widetilde{\varepsilon})$ and
$y_2(\widetilde{t_b}) = 0$, so that object 2 is found left from object 1
on the $y_1$-axis, as for the binary lens.
The value of $\xi \in [0,2\pi)$ for a given $t$ is obtained from (see
Eq.~(\ref{getxi}))
\begin{eqnarray}
2\pi\left(\frac{t - \widetilde{t_b}}{\widetilde{T}} - 
\left\lfloor\frac{t - \widetilde{t_b}}{\widetilde{T}} + \frac{1}{2\pi} 
\left(\widetilde{\xi_0} - \widetilde{\varepsilon} \sin \widetilde{\xi_0}
\right)\right\rfloor\right)\,+ \nonumber \\+\, \widetilde{\xi_0} - 
\widetilde{\varepsilon}\sin \widetilde{\xi_0}
= \xi - \widetilde{\varepsilon} \sin \xi\,.
\end{eqnarray}

The distance of the source objects from the projected lens position 
is given by
\begin{equation}
u_1(t) = \left|{\vec y}^{\,\mathrm{(lens)}}(t) - {\vec y}^{\,(1)}(t)\right|
\end{equation}
and
\begin{equation}
u_2(t) = \left|{\vec y}^{\,\mathrm{(lens)}}(t) - {\vec y}^{\,(2)}(t)\right|\,.
\end{equation}
For a point-mass lens these values can be directly inserted into the
expression for the magnification $A$ of a point-mass lens
(see e.g. Paczy{\'n}ski \cite{pac1}; Griest \& Hu \cite{griesthu})
\begin{equation}
A(u) = \frac{u^2+2}{u\sqrt{u^2+4}}\,.
\end{equation}

Due to the absence of extended caustics, the effect of a rotating binary
source and a point-mass lens is less dramatic than for a 
rotating binary lens. Examples are shown in
Fig.~\ref{rosrcf}.
The parameters have been chosen, so that one gets a binary source
fit for OGLE\#5 for $\widetilde{T} \to \infty$ as trans-configuration
\footnote{The cis-trans-symmetry for binary sources has been discussed
by Dominik \& Hirshfeld (\cite{domhirsh}).}.
This means that $t_\mathrm{E} = 26.27$, 
$\widetilde{\alpha} = 1.4512$, $\widetilde{t_b} = 825.719~\mbox{d}$, 
$\widetilde{b}
= 0.4624$, $\omega = 0.3819$, and $m_\mathrm{base} = -17.9576$.
$\widetilde{\rho} = 0.8679$ has been chosen so that $\widetilde{\rho}$
is the distance in the static case and $\widetilde{m}_1 = 0.5$.
$\widetilde{\beta}$, $\widetilde{\gamma}$, $\widetilde{\varepsilon}$,
and $\widetilde{\xi_0}$ have been chosen as zero and $\widetilde{T}$
takes the values 100~d, 50~d and 25~d.

\begin{figure*}
\vspace{10.5cm}
\caption{Rotating binary sources with parameters as for a binary source
fit for OGLE\#5 (trans-configuration).
$\widetilde{\beta} = \widetilde{\gamma} = 
\widetilde{\varepsilon} = \widetilde{\xi_0} = 0$, and
$\widetilde{m}_1 = 0.5$. {\bf a} $T=100$~d, {\bf b} $T=50$~d, {\bf c} $T=25$~d.}
\label{rosrcf}
\end{figure*}

\section{The parallax effect}
\subsection{Parameters}
The annual motion of the observer (on the earth) around the sun 
gives another effect of rotating binaries. 
It has been mentioned by Gould (\cite{gould}) and observed by
the MACHO collaboration (Alcock et al. \cite{alcock}). 
In contrast to the other cases
of a rotating lens or a rotating source, one knows most of the parameters
of the binary system:
\begin{itemize}
\item the rotation period $T$,
\item the semimajor axis $a_{\oplus}$,
\item the eccentricity $\varepsilon$,
\item the point of time when the earth is in perihelion $t_\mathrm{p}$.
\end{itemize}
One also knows the position of the source
of light characterized by
\begin{itemize}
\item the longitude $\varphi$ measured in the ecliptic plane from the
perihelion towards the earth's motion,
\item the latitude $\chi$ measured from the ecliptic plane towards the
ecliptic north.
\end{itemize}
A fit to an observed light curve for a parallax event involves 
two additional fit parameters:
\begin{itemize}
\item the length of the semimajor axis projected to the lens plane measured
in Einstein radii $\rho'$,
\item a rotation angle $\psi$ in the lens plane describing the relative
orientation of $\vec v_{\perp}$ to the sun-earth system.
\end{itemize}

A displacement of the observer's
position by $\vec{\delta}_\mathrm{O}$ is equivalent to a displacement of the
source position projected to the lens plane by 
\begin{equation}
\vec{\delta}_\mathrm{L} 
= \frac{D_\mathrm{ds}}{D_\mathrm{s}}\,\vec{\delta}_\mathrm{O} = (1-x)\,\vec{\delta}_\mathrm{O}\,.
\end{equation}
If one chooses $x$ and $y$ in the ecliptic plane and $z$ towards the
ecliptic north, where the sun is in the
origin, positive $x$ is into the direction of the perihelion and positive $y$
is from the perihelion towards the earth's motion, the motion of the earth is
given by 
\begin{eqnarray}
x(\xi(t)) & = & a_{\oplus} (\cos(\xi(t))-\varepsilon)\,, \\
y(\xi(t)) & = & a_{\oplus} \sqrt{1-\varepsilon^2} \sin(\xi(t))\,,
\end{eqnarray}
where $\xi(t) \in [0,2\pi)$ can be obtained from
\begin{equation}
2\pi \left(\frac{t - t_\mathrm{p}}{T}-
\left\lfloor\frac{t - t_\mathrm{p}}{T}\right\rfloor\right) = 
\xi - \varepsilon \sin\xi\,.
\end{equation}
This motion has to be projected to the lens plane which is towards the
longitude $\varphi$ and the latitude $\chi$ as defined before.
Let the dimensionless 
coordinates in the lens plane be $\widetilde{x}_1$, $\widetilde{x}_2$,
and let $\widetilde{x}_3$ be a coordinate perpendicular to the lens plane towards
the observer, so that one gets a right-handed system. If one
chooses $\widetilde{x}_1 = z$, $\widetilde{x}_2 = y$ and
$\widetilde{x}_3 = -x$ for $\varphi = \chi = 0$, the angle $\varphi$ gives a
rotation around the $\widetilde{x}_1$-axis and the angle $\chi$ a
following rotation around the $\widetilde{x}_2$-axis, so that one gets
\begin{equation}
\left(\begin{array}{c} \widetilde{x}_1 \\ \widetilde{x}_2 \end{array}
\right) = \frac{1-x}{r_\mathrm{E}}\,\mathcal{R}_2\,
\left(\begin{array}{c}x(\xi(t)) \\
y(\xi(t))\end{array}\right) \label{rotationpar}
\end{equation}
with
\begin{equation}
\mathcal{R}_2 = \left(\begin{array}{cc} -\sin\chi \cos\varphi & -\sin\chi\sin\varphi \\
-\sin \varphi & \cos \varphi\end{array} \right)
\end{equation}
and therefore, with
\begin{equation}
\rho' = \frac{a_{\oplus}(1-x)}{r_\mathrm{E}}\,,
\label{rhosdef}
\end{equation}
one obtains
\begin{eqnarray}
\widetilde{x}_1(t) & = & \rho'\Big[-\sin\chi\cos\varphi(\cos\xi(t)-\varepsilon)
\,- \nonumber\\ & & \left. -\,\sin\chi\sin\varphi\sqrt{1-\varepsilon^2}\sin\xi(t)\right]\,, \\
\widetilde{x}_2(t) & = & \rho'\Big[-\sin\varphi(\cos\xi(t)-\varepsilon)
\,+ \nonumber\\ & & \left. +\,\cos\varphi\sqrt{1-\varepsilon^2}\sin\xi(t)\right]\,.
\end{eqnarray}
The rotation around $\widetilde{x}_3$ by $\psi$ finally gives
\begin{eqnarray}
x_1(t) & = & \cos\psi\,\widetilde{x}_1(t)+\sin\psi\,\widetilde{x}_2(t)\,,\\
x_2(t) & = & -\sin\psi\,\widetilde{x}_1(t)+\cos\psi\,\widetilde{x}_2(t)\,.
\end{eqnarray}
Let $p$ be the parameter along the source trajectory and $d$ the distance
perpendicular to it measured from a line parallel to the
source trajectory through the origin.
By choosing $v_{\perp}$ along $x_1$, I obtain
\begin{eqnarray}
p(t) & = & \widetilde{p_0}(t)+\cos\psi\,\widetilde{x}_1(t)+
\sin\psi\,\widetilde{x}_2(t)\,,\\
d(t) & = & \widetilde{d_0}-\sin\psi\,\widetilde{x}_1(t)\,+
\cos\psi\,\widetilde{x}_2(t)\,,
\end{eqnarray}
where
\begin{equation}
\widetilde{p_0}(t) = \frac{t-\widetilde{t_\mathrm{max}}}{t_\mathrm{E}}\,.
\end{equation}
One sees that for $t = \widetilde{t_\mathrm{max}}$, in general, $p(t) \neq 0$ and
$d(t) \neq \widetilde{d_0}$. To avoid a change in the fit parameters when
including the parallax (i.e. changing between heliocentric and geocentric
coordinates), it is favourable to use the same fit parameters $t_\mathrm{max}$ and
$d_0$ (which has been called $u_\mathrm{min}$ before) in both cases by subtracting
the earth-sun distance at $t_\mathrm{max}$, which yields 
for the coordinates $p(t)$ 
towards the direction of the source
and $d(t)$ perpendicular to it 
\begin{eqnarray}
p(t) & = & p_0(t)+\cos\psi\left(\widetilde{x}_1(t) 
-\,\widetilde{x}_2(t_\mathrm{max})\right) \,+ \nonumber \\ & & +\,\sin\psi
\left(\widetilde{x}_2(t)-\widetilde{x}_2(t_\mathrm{max})\right)\,,\\
d(t) & = & d_0-\sin\psi\left(\widetilde{x}_1(t)
-\widetilde{x}_1(t_\mathrm{max})\right) \,+ \nonumber \\ & & 
+\,\cos\psi\left(\widetilde{x}_2(t)-\widetilde{x}_2(t_\mathrm{max})\right)\,,
\end{eqnarray}
where 
\begin{equation}
p_0(t) = \frac{t-t_\mathrm{max}}{t_\mathrm{E}}\,,
\end{equation}
and the impact parameter is given by
\begin{equation}
u(t) = \sqrt{[d(t)]^2+[p(t)]^2}\,.
\end{equation}

The longitude $\varphi$ and the latitude $\chi$ are related to the ecliptic
coordinates $\beta$ and $\lambda$ in the following way.
The ecliptic coordinates are geocentric
but above a heliocentric system has been used. 
Therefore, the sun-around-earth
motion has to be converted to an earth-around-sun motion. It can be seen
that the vector $\vec x$ of the sun's position measured from the earth
is transformed into a vector $-\vec x$ of the earth's position measured
from the sun. Since the angular momentum $\vec L$ is an axial vector 
(which means that it does not change its sign under this transformation),
the earth moves around the sun in the {\em same} direction as the sun moves
around the earth in a geocentric system. Therefore, one sees that
$\chi = \beta$,
where the parallax is neglected, which does not play a role 
in determining the position, because we
deal with distances of the order of 10~\mbox{kpc}.
The ecliptical length is measured from the vernal equinox along the ecliptic
in the same sense as the right ascension. Since the sun moves towards positive
right ascension,
$\lambda$ increases with time $t$. The earth's motion 
around the sun is also in the direction of positive $\lambda$, so that
$\varphi = \lambda + \varphi_\mathrm{c}$ with a constant $\varphi_\mathrm{c}$, 
if one neglects the
earth-sun distance. The sun's position as seen from the earth corresponds to
$\lambda = 0$ at vernal equinox, while the earth's position as seen from the sun corresponds to
$\lambda = \pi$. If $\varphi_{\gamma}$ denotes the longitude of the vernal equinox
as measured from the perihelion, one obtains 
$\varphi_\mathrm{c} = \pi + \varphi_{\gamma}$ and
therefore 
$\varphi = \lambda + \pi + \varphi_{\gamma}$.

Inserting the definition of the Einstein radius, Eq.~(\ref{Einsteinradius}),
into the definition of $\rho'$, Eq.~(\ref{rhosdef}),
yields
\begin{equation}
\rho ' = \left(\frac{M}{M_{\odot}}\right)^{-1/2}\,
\sqrt{\frac{a_{\oplus}^2}{2 {R_\mathrm{S}}_{\odot} D_\mathrm{s}}\,\frac{1-x}{x}}
\end{equation}
for $x \neq 1$.
One sees that $\rho'$ diverges for $x \to 0$ and $\rho' \propto 1/\sqrt{M}$.
 For $x = \frac{1}{2}$,
$D_\mathrm{s} = 8~\mbox{kpc}$ and $M = 1 M_{\odot}$, $\rho' = 0.7$, while
for $M = 10^{-3}~M_{\odot}$, $\rho' = 23$.

In Fig.~\ref{lcparallax}, a light curve where the earth's motion around the
sun has been considered together with a light curve where this motion has 
been neglected is shown. Since both models use the same parameters, the 
amplification for $t= t_\mathrm{max} = 0$ is the same.
For this example, parameters which are similar to those of
 the parallax event found
by the MACHO collaboration (Alcock et al. \cite{alcock}) have been chosen. 
The eccentricity of the earth's
orbit is $\varepsilon = 0.0167$, while its rotation period is 
$T = 365.26~\mbox{d}$. With the ecliptical coordinates 
$\lambda = 271^{\circ}$ and $\beta = -5^{\circ}$, and
$\varphi_{\gamma} = 77^{\circ}$ being the
longitude of the vernal equinox measured from the perihelion, one obtains
$\varphi = 2.93~\mbox{rad}$ and $\chi = -0.08~\mbox{rad}$. 
Moreover, I have chosen $\psi = 4.14~\mbox{rad}$,
$\rho' = 0.2$, $d_0 = -0.16$, $t_\mathrm{E} = 110~\mbox{d}$, and
$t_\mathrm{max} = t_\mathrm{p} = 0$. 

\begin{figure}
\vspace{5cm}
\caption{Light curves with and without considering the earth's motion
around the sun, the symmetric curve neglects the earth's motion}
\label{lcparallax}
\end{figure}

\section{Additional constraints on the mass and other physical quantities}
\label{estpar:masscon}
For a microlensing event, the Einstein radius $r_\mathrm{E}$, the lens
mass $M$, the lens distance $D_\mathrm{d}$ and the transverse velocity
$v_{\perp}$ cannot be observed directly in general.
Any model for the lens and the source involves the timescale $t_\mathrm{E}$,
which gives the relation
\begin{equation}
r_\mathrm{E} = t_\mathrm{E}\,v_{\perp}\,, \label{rel1}
\end{equation}
so that
$r_\mathrm{E}$ can be eliminated and 3~unknown quantities $M$, $D_\mathrm{d}$
and $v_{\perp}$, or alternatively $\mu = M/M_{\sun}$, $x = D_\mathrm{d}/D_\mathrm{s}$
and $v_{\perp}$ remain, which are related by the definition of the
Einstein radius, which reads
\begin{equation}
t_\mathrm{E}\,v_{\perp} = \sqrt{\frac{4GM_{\sun}}{c^2}\,D_\mathrm{s}}\, \label{rel2}
\sqrt{\mu\,x(1-x)}\,.
\end{equation} 

Additional constraints may arise from certain models of the lens system.
In the following, I discuss constraints from the finite source size, a
rotating binary lens, and the parallax effect.
Combining two of these
allows to determine the lens distance (up to a possible degeneracy) 
and from this value the mass, the velocity, and the Einstein radius.
Using three or more constraints will overdetermine the problem.
However, one should note that there are uncertainties in the fit parameters.

\subsection{Using one constraint}

\subsubsection{From extended sources}
From the fit of an extended source, one obtains an additional constraint
if the physical radius of the source $r_\mathrm{s}$ is known,
which may be obtained approximately using
the color and the absolute magnitude of the source.
The parameter $R_\mathrm{src}$ is the ratio of the physical
radius and the projected Einstein radius $r_\mathrm{E}'$, i.e.
\begin{equation}
R_\mathrm{src} = \frac{r_\mathrm{s}}{r_\mathrm{E}'} = \frac{x r_\mathrm{s}}{r_\mathrm{E}}
= \frac{x r_\mathrm{s}}{t_\mathrm{E}\,v_{\perp}}\,,
\end{equation}
which is an additional constraint between $v_{\perp}$ and $x$, i.e.
\begin{equation}
v_{\perp} (x) = \frac{x r_\mathrm{s}}{t_\mathrm{E}\,R_\mathrm{src}}\,.
\label{con1} \label{vext}
\end{equation}
Since $x \in (0,1)$, one obtains a limit for $v_\perp$:
\begin{equation}
v_\perp < \frac{r_\mathrm{s}}{t_\mathrm{E} R_\mathrm{src}}\,.
\end{equation}
Using the constraint of Eq.~(\ref{con1}), the lens mass can be written as a function of $x$
\begin{equation}
\mu(x) = \frac{c^2}{4GM_{\sun}D_\mathrm{s}}\,\frac{r_\mathrm{s}^2}{R_\mathrm{src}^2}\,\frac{x}{1-x}\,,
\label{massextended}
\end{equation}
or as a function of $v_{\perp}$
\begin{equation}
\mu(v_{\perp}) = \frac{c^2}{4GM_{\sun}D_\mathrm{s}}\,
\frac{t_\mathrm{E} v_{\perp} r_\mathrm{s}^2}
{R_\mathrm{src} (r_\mathrm{s}-t_\mathrm{E} v_{\perp} R_\mathrm{src})}\,.
\end{equation}

\subsubsection{From the observer's motion around the sun}
If one takes into account the observer's motion around the sun,
one has the additional parameter $\rho'$, which is related to the
earth-sun distance $a_{\oplus}$ by
\begin{equation}
\rho' = \frac{a_{\oplus}(1-x)}{r_\mathrm{E}} =
\frac{a_{\oplus}(1-x)}{t_\mathrm{E} v_{\perp}}\,,
\end{equation}
giving a constraint between $x$ and $v_{\perp}$, i.e.
\begin{equation}
v_{\perp}(x) = \frac{a_{\oplus}(1-x)}{t_\mathrm{E} \rho'}\,. \label{vpar}
\end{equation}
Since $x \in (0,1)$, one has
\begin{equation}
v_\perp < \frac{a_\oplus}{t_\mathrm{E} \rho'}\,.
\end{equation}
The mass as a function of $x$ follows as (compare Alcock et al. \cite{alcock})
\begin{equation}
\mu(x) =  \frac{c^2}{4GM_{\sun}D_\mathrm{s}}\,\frac{a_{\oplus}^2}{{\rho'}^2}\,\frac{1-x}{x}\,,
\label{massparallax}
\end{equation}
and as a function of $v_{\perp}$ as
\begin{equation}
\mu(v_{\perp}) =  \frac{c^2}{4GM_{\sun}D_\mathrm{s}}\,\frac{t_\mathrm{E} v_{\perp} a_{\oplus}^2}
{\rho' (a_{\oplus}-t_\mathrm{E} v_{\perp} \rho')}\,.
\end{equation}

\subsubsection{From a rotating binary lens}
For a rotating binary lens, one obtains from the orbital motion
\begin{equation}
T = 2\pi\,\sqrt{\frac{a^3}{GM}} 
= 2\pi t_\mathrm{E}^{3/2} v_{\perp}^{3/2} \sqrt{\frac{\rho^3}{GM_{\sun} \mu}}\,.
\end{equation}
This yields 
the following pairwise relations:
\begin{eqnarray}
\mu(v_{\perp}) &=& \frac{4 \pi^2 t_\mathrm{E}^3 \rho^3}
{G M_{\sun} T^2}\,v_{\perp}^3\,, \label{massvc}\\
\mu(x) &=& \frac{T^4 c^6}{1024 \pi^4 G M_{\sun} \rho^6}
\,\frac{1}{D_\mathrm{s}^3\,x^3 (1-x)^3}\,, \label{massrobin} \\
v_{\perp}(x) &=& \frac{T^2 c^2}{16 \pi^2 t_\mathrm{E} \rho^3}\,
\frac{1}{D_\mathrm{s}\,x(1-x)}\,. \label{vrob}
\end{eqnarray}
Since $x \in (0,1)$, $v_\perp$ is restricted to
\begin{equation}
v_\perp \geq \frac{T^2 c^2}{4 \pi^2 t_\mathrm{E} \rho^3 D_\mathrm{s}}\,.
\end{equation}

\subsection{Using two constraints}

\subsubsection{From an extended source
and the observer's motion around the sun}
Setting equal the expressions for the mass as a function of $x$ for the
extended source (Eq.~(\ref{massextended})) and for the observer's motion
around the sun (Eq.~(\ref{massparallax})), yields
\begin{equation}
\frac{x^2}{(1-x)^2} = \frac{a_{\oplus}^2 R_\mathrm{src}^2}{{\rho'}^2 r_\mathrm{s}^2}\,.
\end{equation}
Since one has
\begin{equation}
\frac{x}{1-x} > 0
\end{equation}
for $x \in (0,1)$, only the positive root is an appropriate solution, and
one obtains
\begin{equation}
\frac{x}{1-x} = \frac{a_{\oplus} R_\mathrm{src}}{{\rho'} r_\mathrm{s}} \equiv X\,.
\end{equation}
Solving for $x$ yields the solution
\begin{equation}
x = \frac{X}{1+X} = \frac{a_\oplus R_\mathrm{src}}{\rho' r_\mathrm{s} +
a_\oplus R_\mathrm{src}}
\end{equation}
and from Eq.~(\ref{vext}) or Eq.~(\ref{vpar}) one obtains for $v_\perp$
\begin{equation}
v_\perp = \frac{a_\oplus r_\mathrm{s}}{t_\mathrm{E} (\rho' r_\mathrm{s}
+ a_\oplus R_\mathrm{src})}\,.
\end{equation}

\subsubsection{From an extended source
and a rotating binary lens}
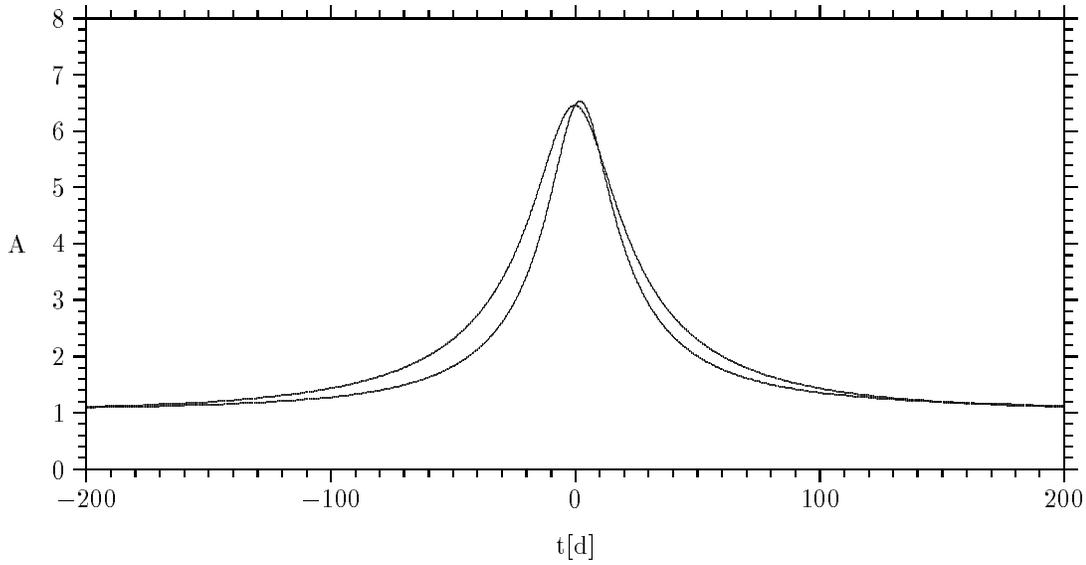
\begin{figure}[htbp]
\vspace{4cm}
\caption{$Y(x)$}
\label{yofx}
\end{figure}

From the  expressions for the mass as a function of $x$ 
(Eq.~(\ref{massextended}) and Eq.~(\ref{massrobin}))
one
obtains the relation
\begin{equation}
(1-x)x^2 = \frac{T^2 c^2 R_\mathrm{src}}{16 \pi^2 \rho^3 D_\mathrm{s} r_\mathrm{s}} \equiv Y\,.
\end{equation}
$Y$ as a function of $x$ 
is shown in Fig.~\ref{yofx}. Note that $x \in (0,1)$.
The function has zeros for the boundary values 
$x=0$ and $x=1$ and a maximum at $(\frac{2}{3},
\frac{4}{27})$. Therefore, $Y$ is restricted to $Y \in (0,\frac{4}{27}]$,
which is a consequence from the constraint on the Einstein radius, which cannot
exceed
\begin{equation}
r_\mathrm{E,max} = \sqrt{\frac{GM D_\mathrm{s}}{c^2}}\,.
\end{equation}
For any value of $Y$, there are two 
values of $x$, except for $Y = Y_\mathrm{max} = \frac{4}{27}$.
For given $x$, the mass $M$, the 
Einstein radius $r_\mathrm{E}$ and the absolute value
of the transverse velocity can be successively calculated.

\subsubsection{From
 the observer's motion around
the sun and a rotating binary lens}

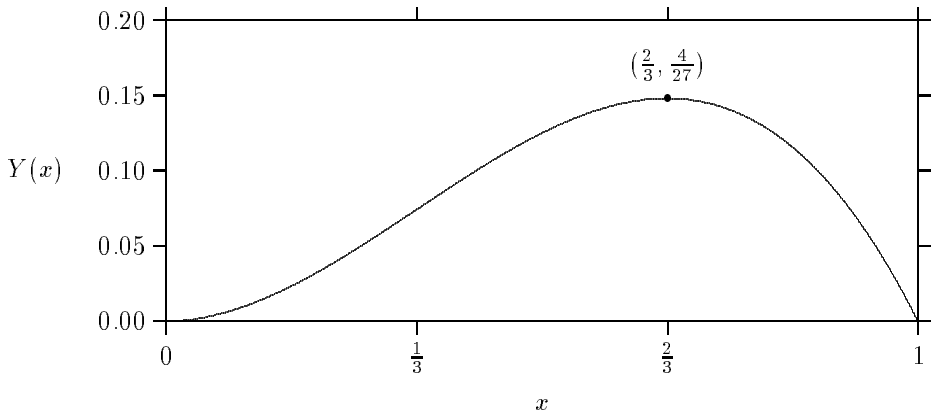
\begin{figure}
\vspace{4cm}
\caption{$Z(x)$}
\label{xiofx}
\end{figure}

From Eq.~(\ref{massparallax}) and Eq.~(\ref{massrobin}) one obtains
\begin{equation}
x(1-x)^2 = \frac{T^2 c^2}{16 \pi^2 D_\mathrm{s} a_{\oplus}}\,\frac{\rho'}{\rho^3}
\equiv Z\,.
\end{equation}
$Z$ as a function of $x$ is shown in Fig.~\ref{xiofx}. Note that
$Z(x) = Y(1-x)$. .
The function has zeros for $x=0$ and $x=1$ and a maximum at $(\frac{1}{3},
\frac{4}{27})$. 
Since $x \in (0,1)$, $Z$ is restricted to $Z \in (0,\frac{4}{27}]$.
For any value of $Z$, there are two 
values of $x$, except for $Z = Z_\mathrm{max} = \frac{4}{27}$.

\subsection{Using three constraints}
With all three constraints, $x$ should be a similar solution to
$X(x)$, $Y(x)$, and $Z(x)$. Since $X(x)$ yields a unique solution, one
of the solutions of $Y(x)$ and $Z(x)$ has to be dropped. Note that
\begin{equation}
X = \frac{Y}{Z}\,.
\end{equation}
Since the fit parameters
and the source radius $r_\mathrm{s}$ contain uncertainties, 
one may fit for a most-likely value of $x$.

\section{A fit with a rotating binary lens for DUO\#2}
\label{DUO2robinfit}
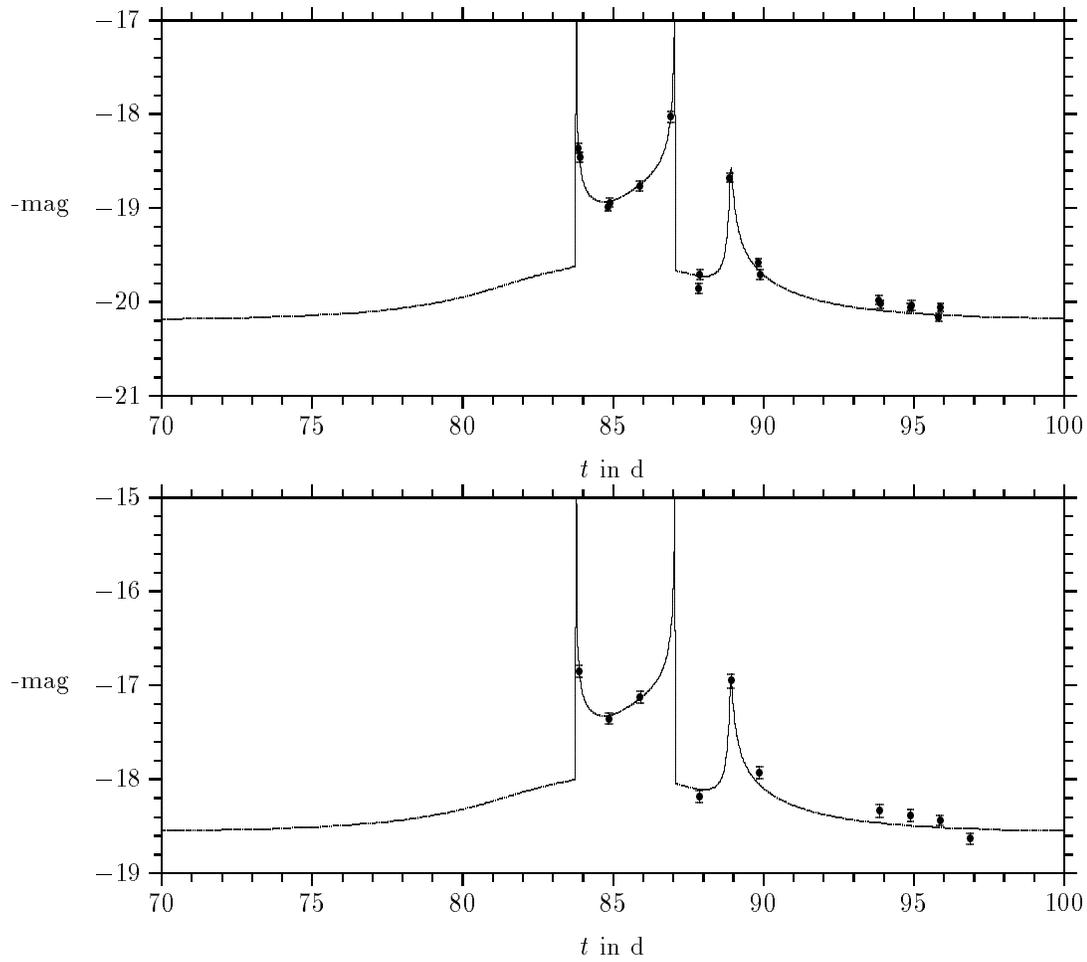
\begin{figure}[htbp]
\vspace{8.5cm}
\caption{DUO \#2: Light curve for a rotating binary lens together
with the data.
Light curve for the blue spectral band on the top and 
light curve for the red spectral band on the bottom.}
\label{d2lcrobin}
\end{figure}

The DUO\#2 event has been reported by Alard et. al (\cite{alcock}), 
where a fit with a strong binary lens is presented. 
I have investigated some more possible fits using
a static binary lens and a point source (Dominik \cite{dominik2}).
Here I show the results of a fit with a rotating binary lens with parameters
which are completely different from the fits with static binaries.
I have omitted one occurence of a data point which
appeared twice in the data I have received from C. Alard, 
so that I use 115 data points and not 116~data
points as in (Alard et al. \cite{alcock}). Moreover, I use the magnitude values for the
fit and not the amplification values.
The parameters for the fit 
are shown in Table~\ref{DUO2robin}, while the
resulting light curves are shown in Fig.~\ref{d2lcrobin}.
The $\chi^2_\mathrm{min}$ indicates that this fit can be marginally accepted.

\begin{table}[htbp]
\caption{Fit for DUO\#2 using a model with a rotating binary lens }
\begin{flushleft}
\begin{tabular}{lc}
\hline\noalign{\smallskip}
parameter & value \\
\noalign{\smallskip}\hline\noalign{\smallskip} 
\rule[-1ex]{0ex}{3.5ex}$t_\mathrm{E}$ [d]& 
6.4 \\
\rule[-1ex]{0ex}{3.5ex}$t_b$ [d]&
85.40
 \\ 
\rule[-1ex]{0ex}{3.5ex}$\rho$ & 
1.058 
 \\ 
\rule[-1ex]{0ex}{3.5ex}$m_1$ &
0.238
 \\ 
\rule[-1ex]{0ex}{3.5ex}$\alpha$ [rad]&
2.374
 \\ 
\rule[-1ex]{0ex}{3.5ex}$b$ & 
-0.087
 \\ 
\rule[-1ex]{0ex}{3.5ex}$m_\mathrm{base,blue}$ &
-20.198
 \\ 
\rule[-1ex]{0ex}{3.5ex}$m_\mathrm{base,red}$ &
-18.564
 \\ 
\rule[-1ex]{0ex}{3.5ex}$f_\mathrm{blue}$ &
0.563
 \\ 
\rule[-1ex]{0ex}{3.5ex}$f_\mathrm{red}$ &
0.547
 \\  
\rule[-1ex]{0ex}{3.5ex}$\beta$ [rad]&
0.958
 \\  
\rule[-1ex]{0ex}{3.5ex}$\gamma$ [rad]&
0.358
 \\  
\rule[-1ex]{0ex}{3.5ex}$T$ [d]&
92.4
 \\  
\rule[-1ex]{0ex}{3.5ex}$\varepsilon$ &
0
 \\  
\rule[-1ex]{0ex}{3.5ex}$\xi_0$ &
-0.046
 \\  
\noalign{\smallskip}\hline\noalign{\smallskip}
\rule[-1ex]{0ex}{3.5ex}$\chi^2_\mathrm{min}$ & 
120.51
 \\ 
\rule[-1ex]{0ex}{3.5ex}$n$ = \# d.o.f. & 
100
 \\ 
\rule[-1ex]{0ex}{3.5ex}$\sqrt{2\chi^2_\mathrm{min}}-\sqrt{2n-1}$ & 
1.418
 \\ 
\rule[-1ex]{0ex}{3.5ex}$P(\chi^2 \geq \chi^2_\mathrm{min})$ & 
8~\%
 \\ \noalign{\smallskip}\hline
\end{tabular}
\end{flushleft}
\label{DUO2robin}
\end{table}

Note that the peak after the second caustic crossing is to some part due to the 
rotation and that $R_T$ is 0.07.

If one adopts a transverse velocity of $v_{\perp} = 30~\mbox{km}/\mbox{s}$,
one obtains, using Eq.~(\ref{massvc}), a total mass of the lens of $M = 0.025~M_{\sun}$,
i.e. one object with $M_1 = 5.9 \cdot 10^{-3}~M_{\sun}$ and one object with mass $M_2 =
0.019~M_{\sun}$. The Einstein radius becomes $r_\mathrm{E} = 0.11~\mbox{AU}$ and
the semimajor axis $a = 0.12~\mbox{AU}$.  
However, the lens distance parameter $x$ follows as $0.993$ or $0.007$,
which seems improbable.
Taking 0.993, i.e. the lens close to the source, as
the more probable configuration, one can understand the lens more easily if it
is in the bulge population rather than in the disk population. However, this
value is very extreme, a value of 0.9 or 0.95 would have been more plausible,
but there is some room in the uncertainty of the velocity\footnote{Note that
$M \sim v_\perp^3$.} and of the
fit parameters. Anyway, I have tried this model mainly to show that there
are parameters for a rotating binary model distinct from the static ones 
which produce the light curve,
without assuming that a physically reasonable model would result. 

As noted in Sect.~5, long timescales $t_\mathrm{E}$ are favoured for showing
rotation effects and the timescale for DUO\#2 is rather short. It is 
interesting to see how the situation changes if an event would have been
observed whose light curve shows the same shape as that of DUO\#2 but
whose timescale is 10 times larger. For this hypothetical event, the only
changes in the fit parameters compared to DUO\#2 
would be an enlargement in $t_\mathrm{E}$ and $T$ by a factor of 10.
For the same transverse velocity $v_\perp = 30~\mbox{km}/\mbox{s}$,
the Einstein radius $r_\mathrm{E}$ would be 10 times larger and,
according to Eqs.~(\ref{massvc}) and~(\ref{vrob}), the mass $M$ and $x(1-x)$ would
increase by the same factor, yielding $M_1 = 0.059~M_{\sun}$, 
$M_2 = 0.19~M_{\sun}$,
$r_\mathrm{E} = 1.1~\mbox{AU}$, $a = 1.2~\mbox{AU}$, and $x = 0.92$,
i.e. more reasonable values, so that the light curve (whose shape is identical
to that of DUO\#2) may well be explained by this rotation effect.
This shows that it is worth trying
fits with rotating binary lenses.

\section{Summary}
Rotating binaries are a reality both in the universe and among galactic 
microlensing observations. They are helpful in providing additional information
about physical parameters of the lens system and they may also be used to
assign probabilities to fits using the knowledge on the distribution of
their parameters.
The inclusion of the rotation for binary lenses enlarges the parameter
space and gives room for additional parameter degeneracies.
It also provides additional shapes of light curves through the motion of the
caustics.
Every fit with a static binary should be checked for consistency.

\begin{acknowledgements}
I would like to thank S.~Mao for some discussions on the subject and for useful
comments, 
C.~Alard for sending me the data of the DUO\#2 event, the OGLE collaboration
for making available their data, and the MACHO collaboration for making available
the MACHO LMC\#1 data.
\end{acknowledgements}

\appendix
\section{Approximation of the trajectory to first order
in $\varepsilon$}
Here I show that the expressions for parallax light curves
given by Alcock et. al (\cite{alcock}) are reproduced by expanding the 
trajectory to first order in the eccentricity $\varepsilon$.
One obtains the earth's trajectory in the form
(e.g. Montenbruck \& Pfleger \cite{mont})
\begin{equation}
\left(\begin{array}{c} x(t) \\ y(t)\end{array}\right) =
A(t)\,\left(\begin{array}{c} \cos \xi(t) \\ \sin \xi(t)\end{array}\right)
\end{equation}
with
\begin{eqnarray}
A(t) & = & a_{\oplus} \left(1-\varepsilon \cos \left(2\pi\,
\frac{t-t_\mathrm{p}}{T}\right)\right)\\
\xi(t) & = & 2\pi\, \frac{t-t_\mathrm{p}}{T} + 2 \varepsilon \sin \left(
2\pi\,\frac{t-t_\mathrm{p}}{T}\right)\,.
\end{eqnarray}
With
\begin{equation}
A'(t) = A(t)\frac{1-x}{r_\mathrm{E}}
\end{equation}
the lens plane coordinates $\widetilde{x}_1(t)$ and
$\widetilde{x}_2(t)$ follow from Eq.~(\ref{rotationpar}) as
\begin{eqnarray}
\widetilde{x}_1(t) & = & A'(t) \left[-\sin\chi\cos\varphi\cos\xi(t) \,-\right. \nonumber \\
&& \left.-\,\sin\chi\sin\varphi\sin\xi(t)\right]\,, \label{x1long} \\
\widetilde{x}_2(t) & = & A'(t) \left[-\sin\varphi\cos\xi(t)+
\cos\varphi\sin\xi(t)\right]\,. \label{x2long}
\end{eqnarray}
For $\varphi = 0$ one has
\begin{eqnarray}
\widetilde{x}_1(t) & = & -A'(t)\sin\chi\cos\xi(t)\,, \label{x1short} \\
\widetilde{x}_2(t) & = & A'(t)\sin\xi(t)\,, \label{x2short}
\end{eqnarray}
and therefore
\begin{eqnarray}
d(t) & = & \widetilde{d_0} + A'(t) \cos\psi \sin\xi(t)\,+ \nonumber \\
& & +\,A'(t)\sin\psi\sin\chi\cos\xi(t)\,,\\
p(t) & = & \widetilde{p_0}(t) - A'(t) \cos\psi \sin\chi\cos\xi(t)\,+ \nonumber \\
& & +\,A'(t) \sin\psi\sin\xi(t)\,.
\end{eqnarray}
For the distance from the origin $u(t)$ one gets
\begin{eqnarray}
u^2(t) & = & [d(t)]^2+[p(t)]^2 = \widetilde{d_0}^2 + \widetilde{p_0}^2 
+ {A'}^2 \sin^2\xi + \nonumber \\
& & + {A'}^2\sin^2\chi\cos^2\xi +\nonumber  \\
& & +2 A'\sin\xi[\widetilde{d_0} \cos\psi+\widetilde{p_0} \sin \psi] + \nonumber \\
& & +2 A'\sin\chi\cos\xi[\widetilde{d_0}\sin\psi-\widetilde{p_0} \cos \psi]\,.
\end{eqnarray}
The rotation in the ecliptic plane by an angle $\varphi$ is equivalent
to a shift in $\xi(t)$. This means that one has
\begin{equation}
\xi(t) = 2\pi\,\frac{t-t_\mathrm{p}}{T} + 2 \varepsilon \sin \left(
2\pi\, \frac{t-t_\mathrm{p}}{T}\right)-\varphi = \xi_0(t)-\varphi\,.
\end{equation}
Inserting $\xi(t)$ into Eqs.~(\ref{x1short}) and~(\ref{x2short}) reveals the
expressions given in Eqs.~(\ref{x1long}) and~(\ref{x2long}) where $\xi(t)$ has to
be replaced by $\xi_0(t)$.

The MACHO collaboration (Alcock et al. \cite{alcock}) find that
\begin{eqnarray}
u^2(t) & = & u_0^2 + \omega^2(t-t_0)^2+\alpha^2\sin^2[\Omega(t-t_\mathrm{c})] \nonumber \\
 & & + 2\alpha\sin[\Omega(t-t_\mathrm{c})][\omega(t-t_0)\sin\theta
+u_0\cos\theta] +\nonumber \\
& & + \alpha^2\sin^2\beta \cos^2[\Omega(t-t_\mathrm{c})] + \nonumber\\
& & + 2 \alpha\sin\beta\cos[\Omega(t-t_\mathrm{c})] \,\cdot \nonumber \\
& & \quad \cdot\,[\omega(t-t_0)\cos\theta
- u_0\sin\theta]\,,
\end{eqnarray}
where
\begin{equation}
\alpha = \frac{\omega a_{\oplus}}{\tilde v}
\left(1-\varepsilon\cos[\Omega_0(t-t_\mathrm{p})]\right)
\end{equation}
and
\begin{equation}
\Omega(t-t_\mathrm{c}) = \Omega_0(t-t_\mathrm{c})+ 2\varepsilon
\sin[\Omega_0(t-t_\mathrm{p})]\,,
\end{equation}
where $\Omega_0 = 2\pi/T$ and $\tilde v = v_{\perp}/(1-x)$.

This expression is equivalent to the derived one with
$\alpha = A'$, $u_0 = \widetilde{d_0}$, $\Omega(t-t_\mathrm{c}) = \xi$, 
$\widetilde{p_0} = -\omega(t-t_0)$, $\theta = -\psi$, $\varphi = \Omega_0(t_\mathrm{c}-t_\mathrm{p})$.
Note the sign in $\widetilde{p_0}$ and $\theta$, which is due to the 
fact that I define the lens to be on the right side of the moving source,
while the MACHO collaboration lets the source move in the
opposite direction. Further note that their replacement of $\varphi$
by $\Omega_0(t_\mathrm{c}-t_\mathrm{p})$ is an approximation and that $t_\mathrm{c}$
is the point of time where the earth is closest to the sun-source line.
Finally note that $\omega = 1/t_\mathrm{E}$ and
\begin{equation}
\rho' = \frac{\omega a_{\oplus}}{\widetilde{v}}
= \frac{a_{\oplus}}{t_\mathrm{E} \widetilde{v}}\,.
\end{equation}

\clearpage

\begin{figure*}
{\LARGE Figure 1a}\\[5mm]

\epsfig{file=h0399.f1a}
\end{figure*}

\begin{figure*}
{\LARGE Figure 1b}\\[5mm]

\epsfig{file=h0399.f1b}
\end{figure*}

\begin{figure*}
{\LARGE Figure 1c}\\[5mm]

\epsfig{file=h0399.f1c}
\end{figure*}

\begin{figure*}
{\LARGE Figure 1d}\\[5mm]

\epsfig{file=h0399.f1d}
\end{figure*}

\begin{figure*}
{\LARGE Figure 2a}\\[5mm]

\epsfig{file=h0399.f2a}
\end{figure*}

\begin{figure*}
{\LARGE Figure 2b}\\[5mm]

\epsfig{file=h0399.f2b}
\end{figure*}

\begin{figure*}
{\LARGE Figure 2c}\\[5mm]

\epsfig{file=h0399.f2c}
\end{figure*}

\begin{figure*}
{\LARGE Figure 2d}\\[5mm]

\epsfig{file=h0399.f2d}
\end{figure*}

\begin{figure*}
{\LARGE Figure 3a}\\[5mm]

\epsfig{file=h0399.f3a}
\end{figure*}

\begin{figure*}
{\LARGE Figure 3b}\\[5mm]

\epsfig{file=h0399.f3b}
\end{figure*}

\begin{figure*}
{\LARGE Figure 4a}\\[5mm]

\epsfig{file=h0399.f4a}
\end{figure*}

\begin{figure*}
{\LARGE Figure 4b}\\[5mm]

\epsfig{file=h0399.f4b}
\end{figure*}

\begin{figure*}
{\LARGE Figure 4c}\\[5mm]

\epsfig{file=h0399.f4c}
\end{figure*}

\begin{figure*}
{\LARGE Figure 5}\\[5mm]

\epsfig{file=h0399.f5}
\end{figure*}

\begin{figure*}
{\LARGE Figure 6}\\[5mm]

\epsfig{file=h0399.f6}
\end{figure*}

\begin{figure*}
{\LARGE Figure 7}\\[5mm]

\epsfig{file=h0399.f7}
\end{figure*}

\begin{figure*}
{\LARGE Figure 8}\\[5mm]

\epsfig{file=h0399.f8a}
\epsfig{file=h0399.f8b}
\end{figure*}


\begin{thebibliography}{}
\bibitem[1995]{alard}
Alard C., Mao S., Guibert J., 1995, A\&A 300, L17
\bibitem[1995]{alcock}
Alcock C., Allsman R. A., Alves D., et al., 1995, ApJ 454, L125
\bibitem[1995]{ansari}
Ansari R., Cavalier F., Couchot F., et al., 1995, A\&A 299, 168
\bibitem[1997a]{dominik1}
Dominik M., 1997a, Estimating physical quantities for an observed
galactic microlensing event, preprint astro-ph/9701035, submitted (D97a)
\bibitem[1997b]{dominik2}
Dominik M., 1997b, Ambiguities in fits of observed galactic microlensing events,
preprint astro-ph/9702039, submitted
\bibitem[1996]{domhirsh}
Dominik M., Hirshfeld A. C., 1996, A\&A 313, 841
\bibitem[1992]{gould}
Gould A., 1992, ApJ 392, 442
\bibitem[1992]{griesthu}
Griest K., Hu W., 1992, ApJ, 397, 362 (Erratum: 1993, ApJ 407, 440)
\bibitem[1969]{landau}
Landau L. D., Lifshitz E. M., 1969, Mechanics, Volume 1 of Course of Theoretical
Physics, 2nd edition, Pergamon Press, Oxford
\bibitem[1989]{mont}
Montenbruck O., Pfleger T., 1989, Astronomie mit dem Personal Computer,
Springer, Berlin
\bibitem[1986]{pac1}
Paczy{\'n}ski B., 1986, ApJ 304, 1
\bibitem[1996]{pac2}
Paczy{\'n}ski B., 1996, ARA\&A 34, 419
\bibitem[1992]{sef}
Schneider P., Ehlers J., Falco E. E., 1992, Gravitational Lenses,
Springer, Berlin


\end{thebibliography}
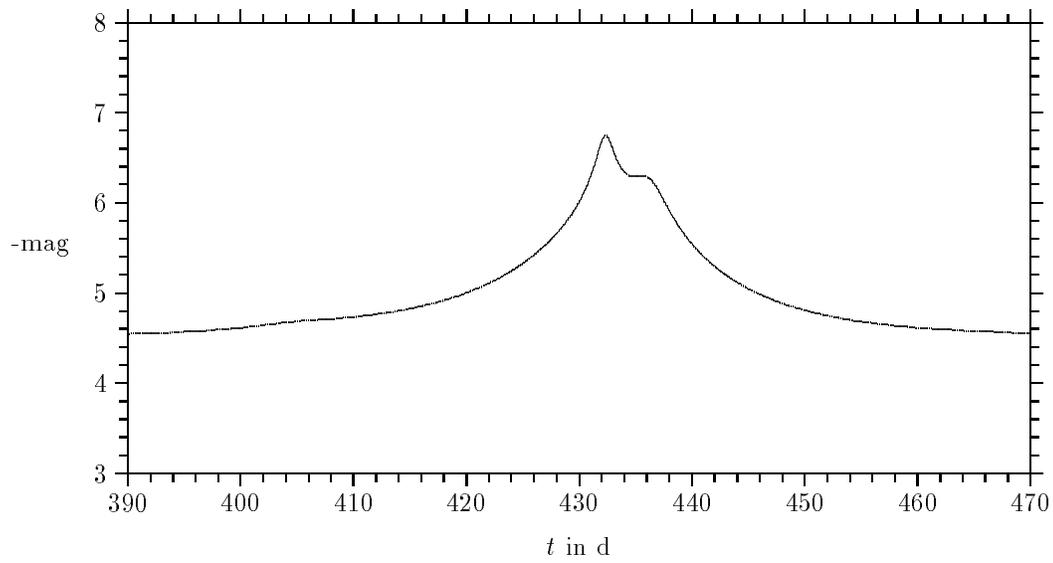
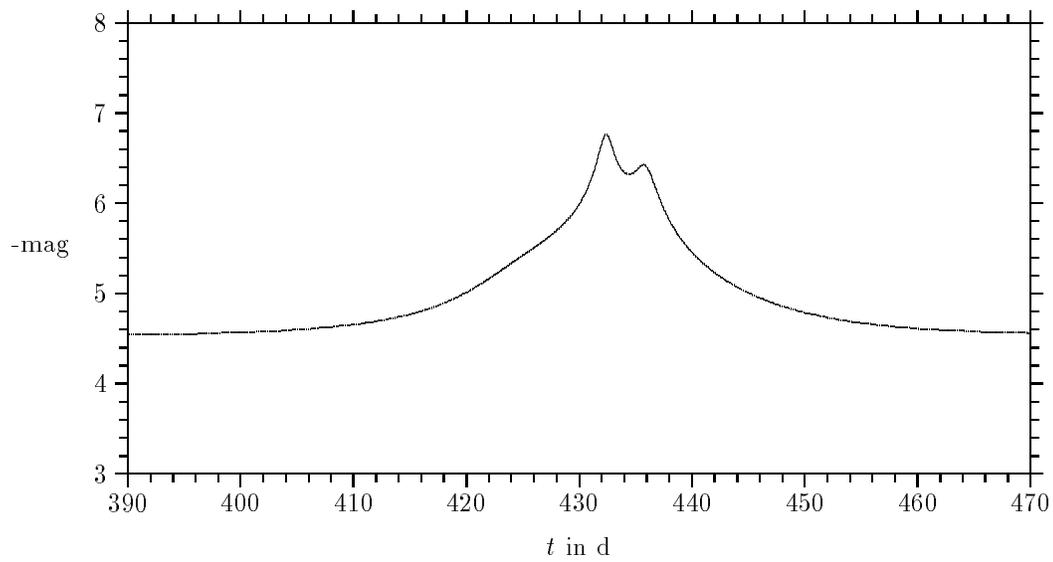
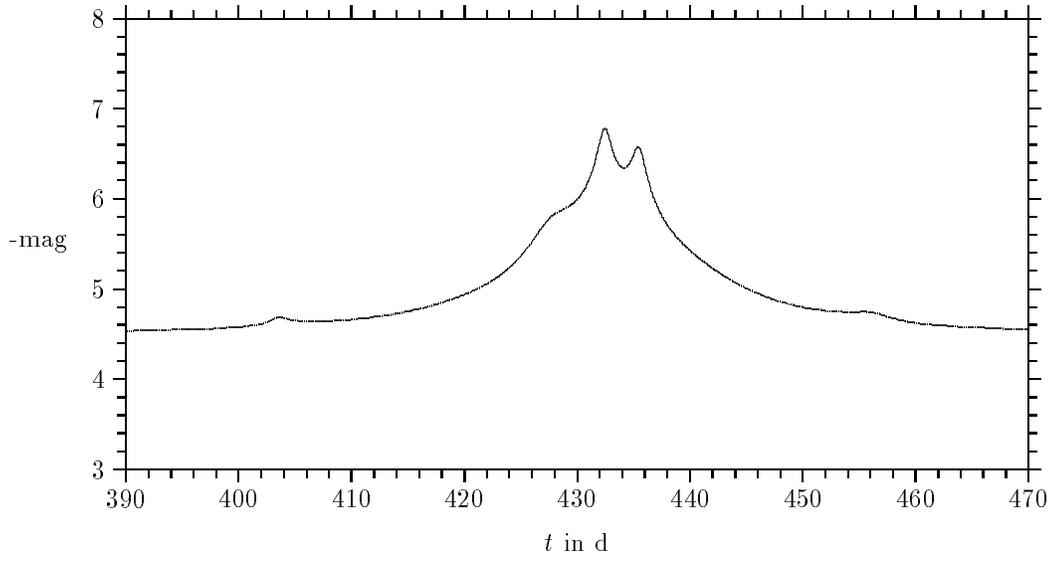
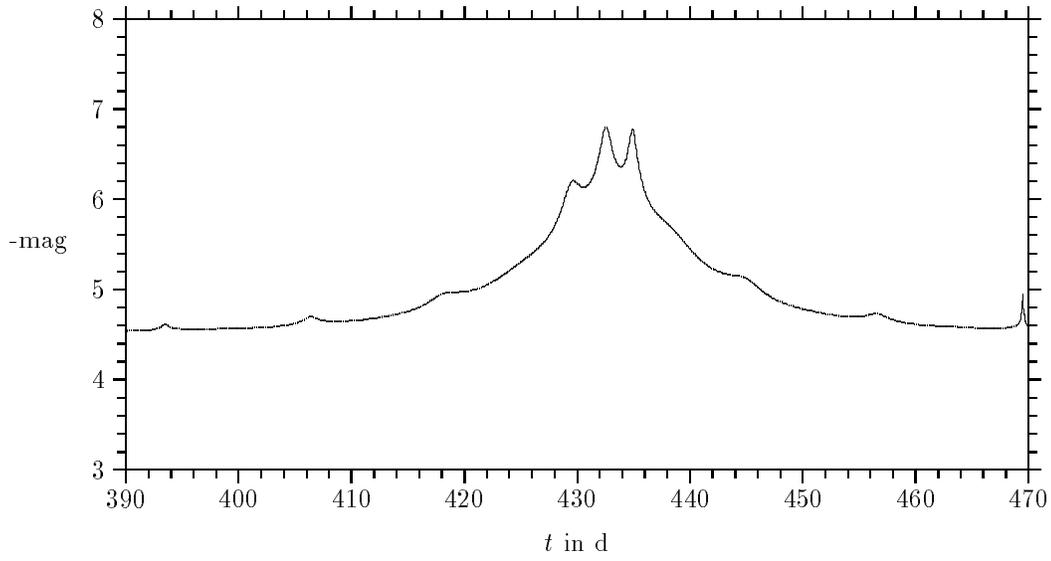
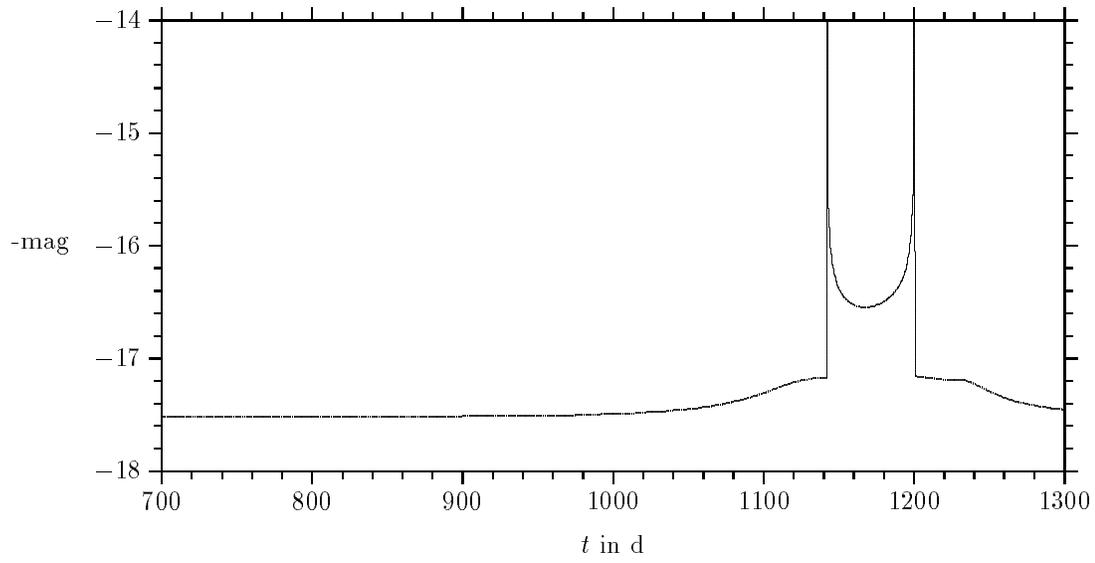
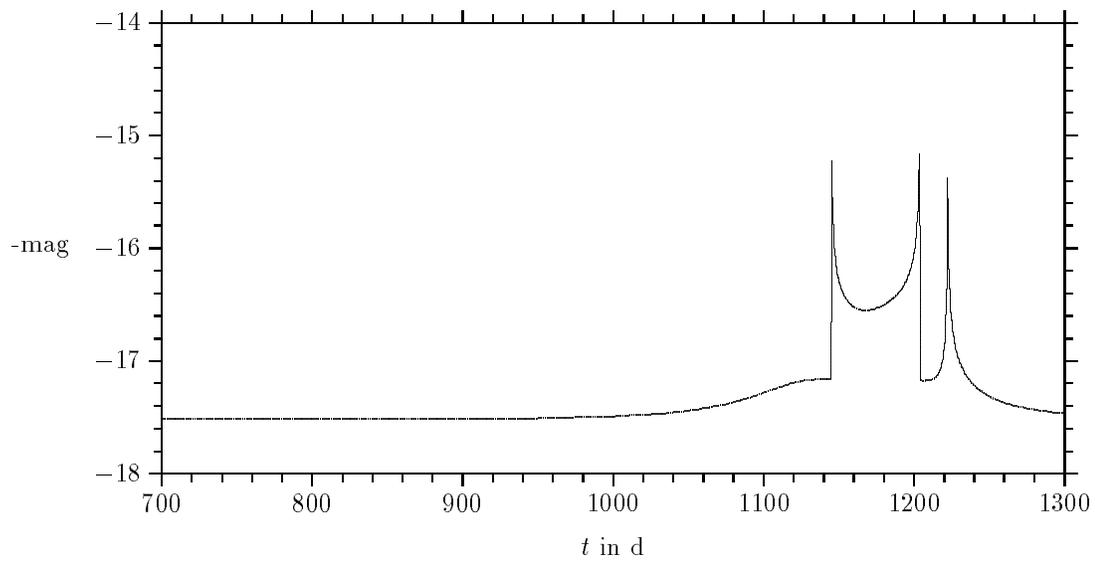
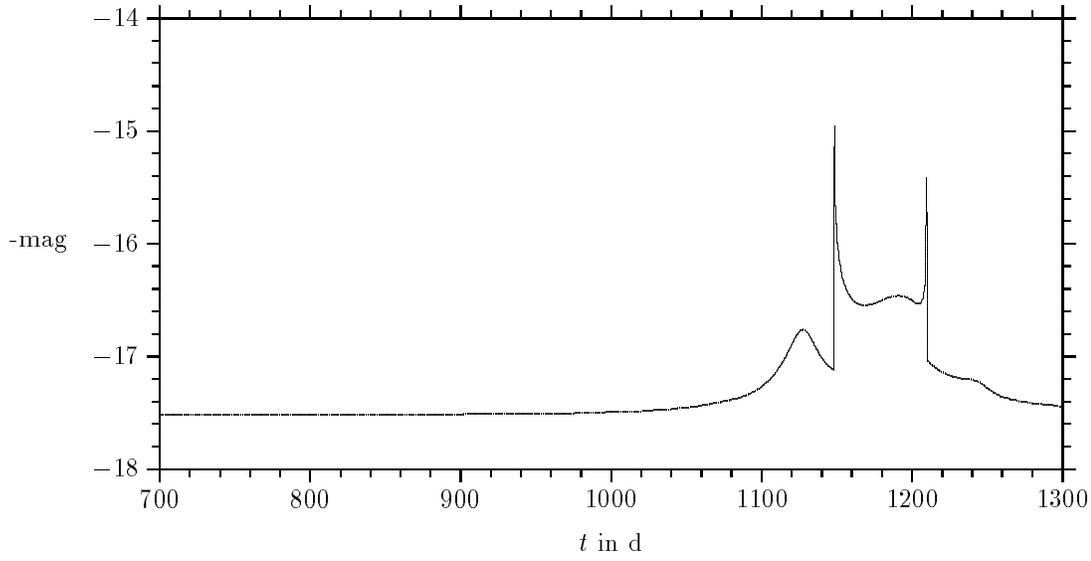
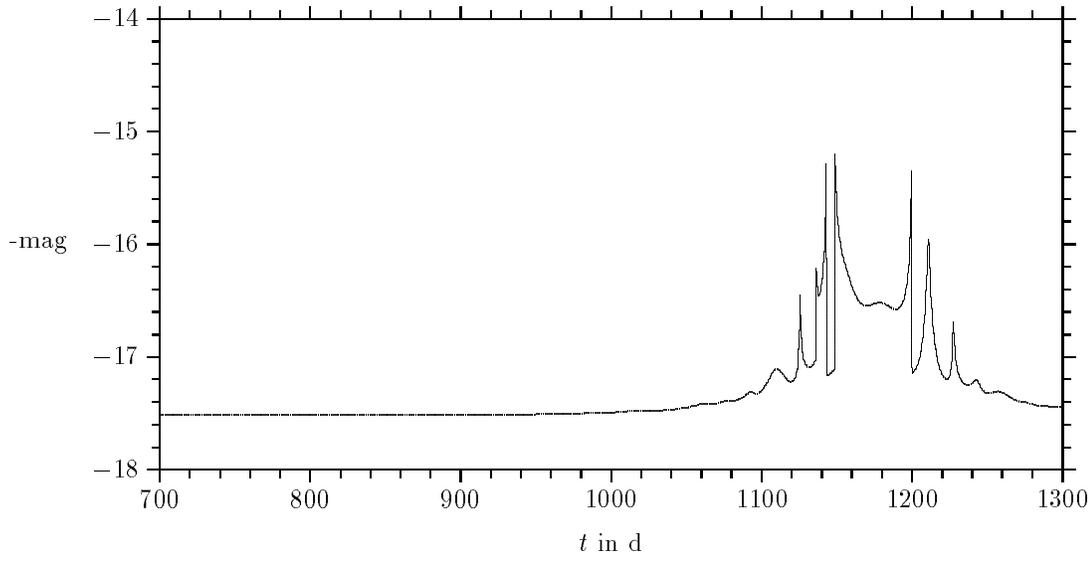
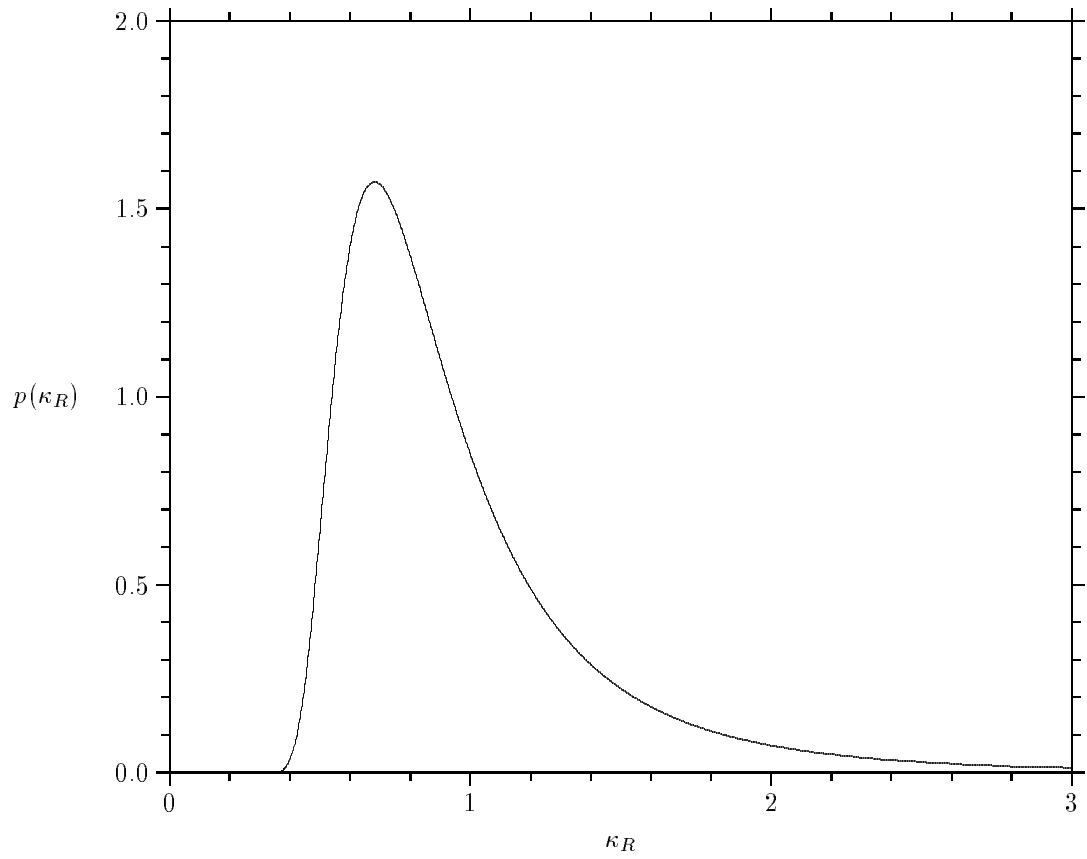
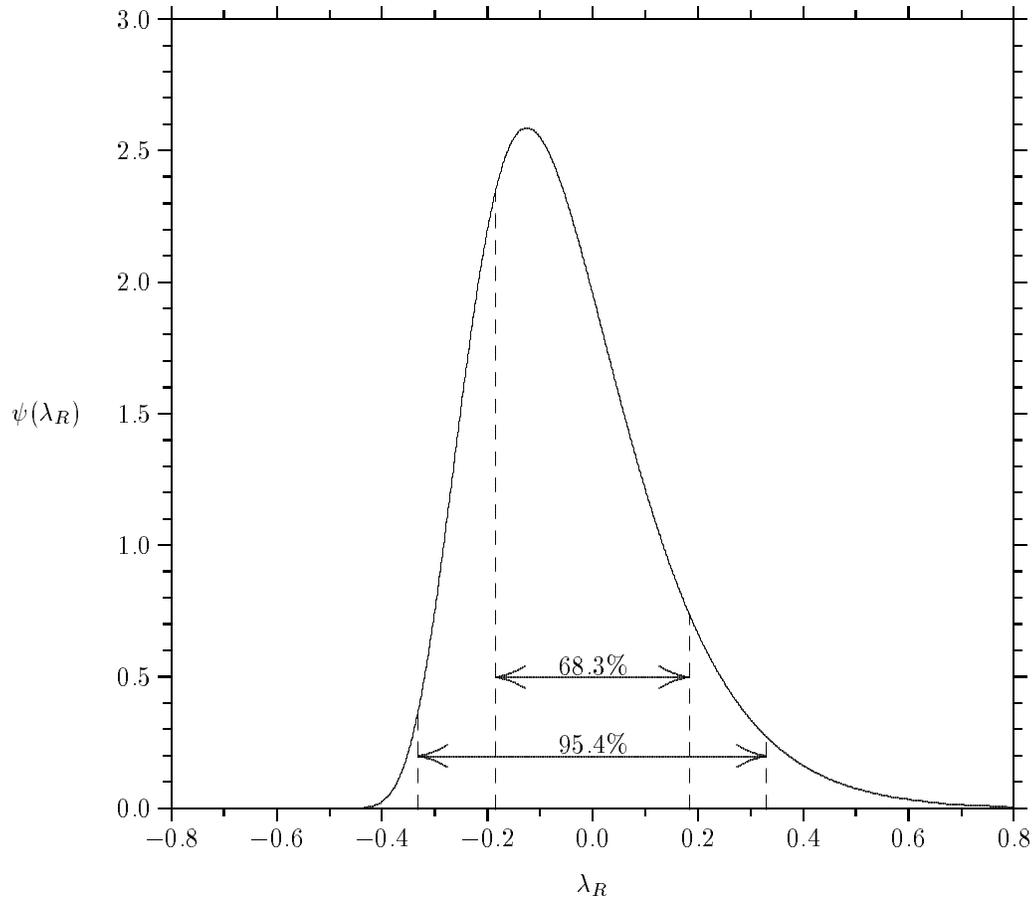
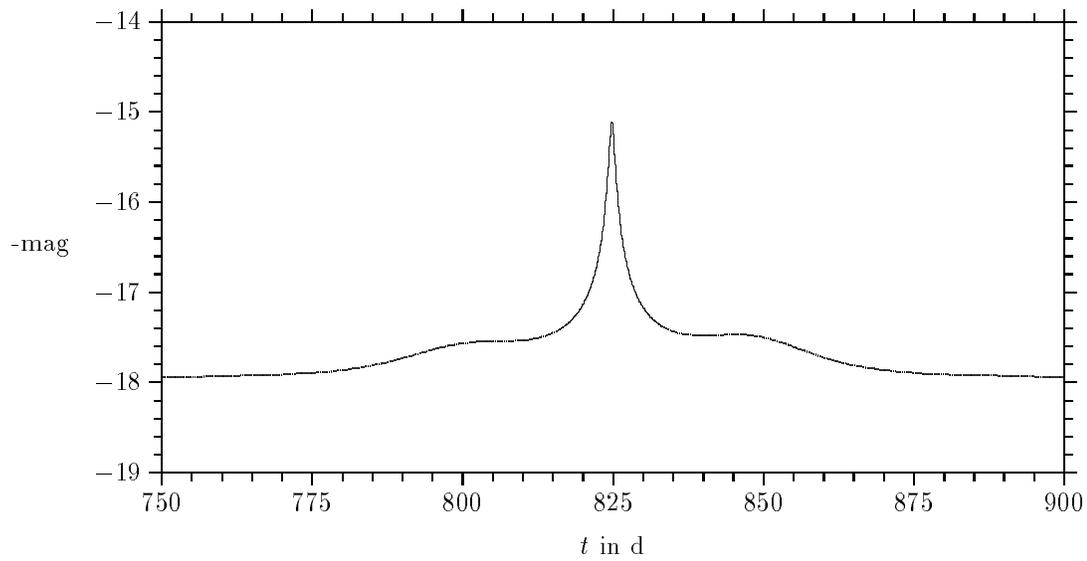
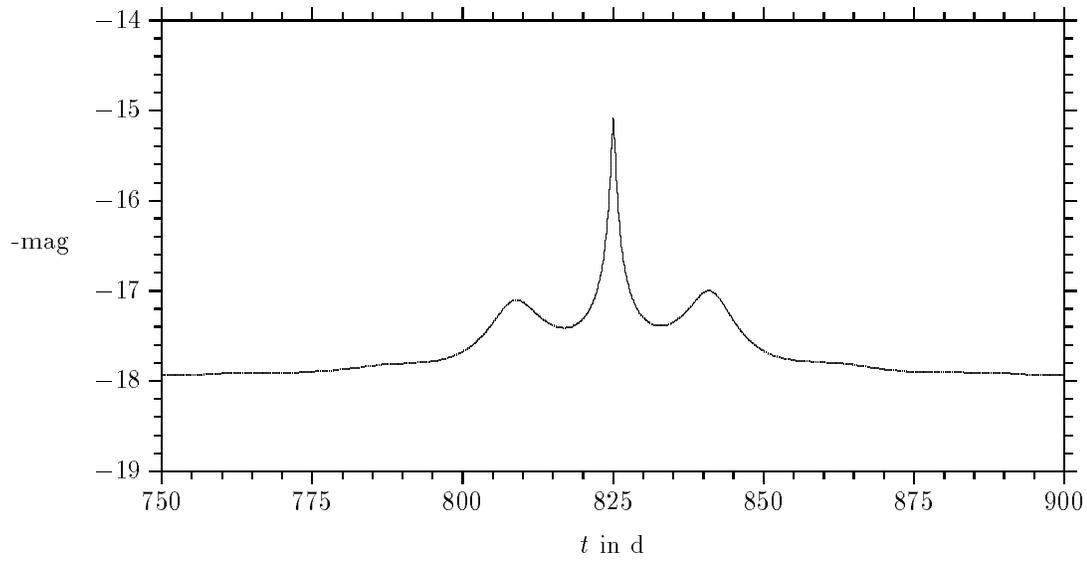
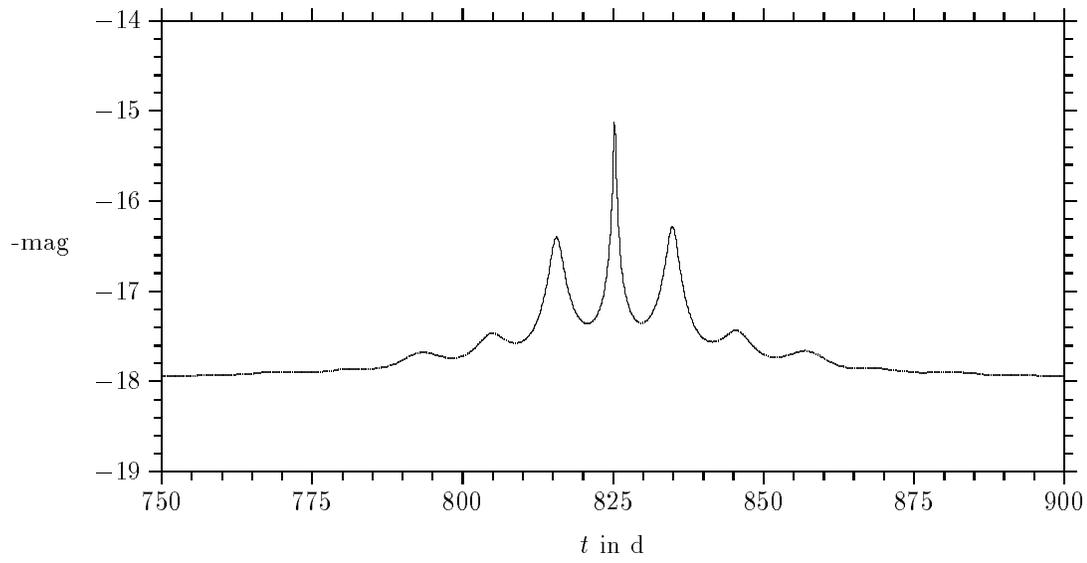
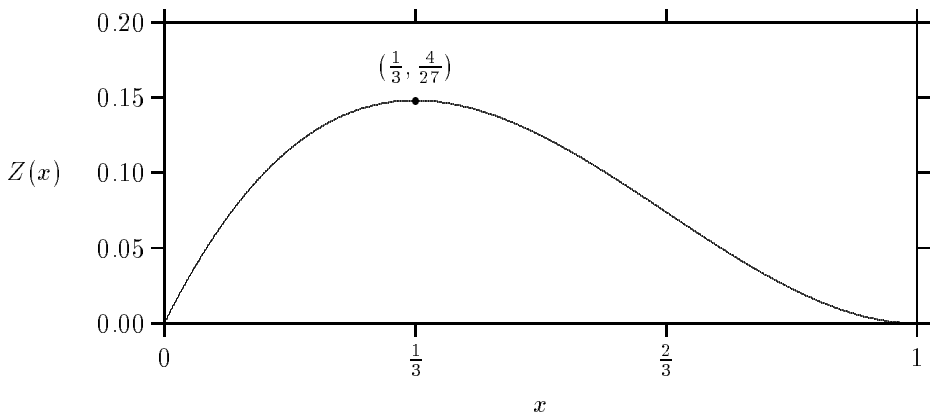
\end{document}